\documentclass[namedreferences]{solarphysics}
\usepackage[hyperref,optionalrh]{spr-sola-addons} 
\usepackage{graphicx}        
\usepackage{amssymb}
\usepackage{savesym}
\savesymbol{iint}
\savesymbol{iiint}
\usepackage{amsmath}
\usepackage{color}           
\usepackage{breakurl}        
\usepackage{lineno, blindtext}

\renewcommand{\vec}[1]{{\mathbfit #1}}

\newcommand{\curl}{ {\bf \nabla} \times}

\chardef\us=`\_

\begin{document}

\begin{article}
\begin{opening}

\title{Two Classes of Eruptive Events during Solar Minimum}

\author[addressref={},corref,email={prantika.bhowmik@durham.ac.uk}]{\inits{P.}\fnm{P.}~\lnm{Bhowmik}\orcid{0000-0002-4409-7284}}
\author[addressref={},email={anthony.yeates@durham.ac.uk}]{\inits{A. R.}\fnm{A.R.}~\lnm{Yeates}\orcid{0000-0002-2728-4053}}

\address[]{Department of Mathematical Sciences, Durham University, Durham, DH1 3LE, UK}

\runningauthor{P. Bhowmik, A.R. Yeates}
\runningtitle{Eruptive Events during Solar Minimum}

\begin{abstract}
During solar minimum, the Sun is relatively inactive with few sunspots observed on the solar surface. Consequently, we observe a smaller number of highly energetic events such as solar flares or coronal mass ejections (CMEs), which are often associated with active regions on the photosphere. Nonetheless, our magnetofrictional simulations during the minimum period suggest that the solar corona is still dynamically evolving in response to the large-scale shearing velocities on the solar surface. The non-potential evolution of the corona leads to the accumulation of magnetic free energy and helicity, which is periodically shed in eruptive events. We find that these events fall into two distinct classes: One set of events are caused by eruption and ejection of low-lying coronal flux ropes, and they could explain the origin of occasional CMEs during solar minimum. The other set of events are not driven by destabilisation of low-lying structures but rather by eruption of overlying sheared arcades. These could be associated with streamer blowouts or stealth CMEs. The two classes differ significantly in the amount of magnetic flux and helicity shed through the outer coronal boundary. We additionally explore how other measurables such as current, open magnetic flux, free energy, coronal holes, and the horizontal component of the magnetic field on the outer model boundary vary during the two classes of event. This study emphasises the importance and necessity of understanding the dynamics of the coronal magnetic field during solar minimum.  
\end{abstract}
\end{opening}

\section{Introduction}

The magnetic-field structure in the solar corona is primarily governed by evolution of the photospheric magnetic field. Large-scale systematic flows and random convective motion of plasma on the solar surface redistribute the magnetic field; consequently, the coronal magnetic field evolves in a time-dependent manner. However, most global computational models assume quasi-static evolution of the coronal magnetic field such that they produce a sequence of independent ``single-time'' extrapolations from the photospheric magnetic fields at discrete time intervals \citep{2012LRSP....9....6M}. The premise is motivated by the relatively fast response time for any disturbance originating in the photosphere to propagate through the corona. The surface magnetic field encounters plasma flow of speed $1\,--\,2$\,km\,s$^{-1}$ (for e.g., differential rotation; \citeauthor{2009LRSP....6....1H}, \citeyear{2009LRSP....6....1H}), whereas any perturbation in the corona is dispersed with the Alfv\'{e}n speed (a few $1000$\,km\,s$^{-1}$ in coronal loops; \citeauthor{2007Sci...317.1192T}, \citeyear{2007Sci...317.1192T}). Thus, the assumption of a quasi-static corona is reasonably justified. Such models have been successful in reproducing various observed coronal magnetic-field features (\citeauthor{2012LRSP....9....6M}, \citeyear{2012LRSP....9....6M} and references therein). Models based on nonlinear force-free extrapolations perform well regarding active-region associated structures, where reliable vector magnetic-field input data are available. However, these static models cannot capture the build-up of free magnetic energy outside of strong active regions, such as in filament channels \citep{2018SSRv..214...99Y}. Because static coronal models -- even those solving the full magneto-hydrodynamics (MHD) equations -- ignore any previous magnetic connectivity that existed in the corona.

In contrast, the modelling approach used in this study preserves the ``memory'' imparted by the slowly evolving photospheric-field distribution over time. The technique was initially introduced by \cite{2000ApJ...539..983V} and was later utilised (with necessary modifications and improvements) in many studies on the evolution of the non-potential coronal magnetic field. The methodology is based on a magnetofrictional approach where the velocity within the corona is proportional to the Lorentz force, such that coronal magnetic field relaxes toward a force-free equilibrium \citep{1986ApJ...309..383Y}. Note that the magnetofriction model is not a dynamic model like a full MHD model, and indeed it acts like a quasi-static model. However, it can capture the effect of magnetic reconnection through the diffusion term present in the magnetic-induction equation (used in this model), which again dictates the dynamic evolution of the magnetic field. In this sense, the magnetofriction model is more advanced than other static non-linear force-free coronal models (for detailed comparison, refer to \citeauthor{2018SSRv..214...99Y}, \citeyear{2018SSRv..214...99Y}). Magnetofriction simulations were successful in explaining the observed hemispheric pattern in solar filaments \citep{2000ApJ...544.1122M,2005ApJ...621L..77M,2006ApJ...641..577M,2008SoPh..247..103Y,2012ApJ...753L..34Y}. These models were also applied to successfully model the formation of non-potential magnetic structures, and associated eruptive phenomena in the global solar corona \citep{2006ApJ...641..577M,2006ApJ...642.1193M,2009ApJ...699.1024Y,2012ApJ...757..147C,2014SoPh..289..631Y,2016ApJ...828...83G,2017ApJ...846..106L}. \cite{2016MNRAS.456.3624G} used a magnetofriction model with different parameter combinations to explore coronal activity in other stars. However, the magnetofrictional model has some limitations too. For example, unlike MHD simulations, it cannot provide information about the plasma properties in the solar corona. Also, the solar wind is implemented in a simplified way: the model uses a radial outflow boundary condition to mimic the effect of the solar wind. Using a full MHD model, \cite{2018NatAs...2..913M} presented a prediction of the global coronal magnetic-field distribution for the August 2017 solar eclipse, which again used a magnetofriction model to inform the energisation of filament channels in the MHD model. From the same MHD model, they were able to deliver the associated coronal density and brightness profiles, which cannot be derived using a magnetofriction model only. 

Most earlier research with the magnetofriction model aimed to study the formation and evolution of non-potential structures such as flux ropes in the solar corona, and their sudden eruption. The same is the objective of our work presented here, but we focus in detail on the evolution of such structures during the solar minimum period. Non-potential structures are generated in the corona by the surface evolution, and form, in general, over polarity-inversion lines as a result of flux cancellation and magnetic reconnection \citep{1989ApJ...343..971V}. They may take the form of sheared arcades or flux ropes where a bundle of magnetic-field lines are twisted around a common axis. Corresponding free magnetic energy typically becomes concentrated above polarity inversion lines, in so-called filament channels \citep{2010SSRv..151..333M}. Naturally, magnetic helicity is also prone to accumulate in these structures \citep{2015ApJ...809..137K}. Magnetic helicity is an important topological quantity with numerous applications in many astrophysical systems \citep{1999PPCF...41B.167B}. In the case of solar corona, it serves as a quantitative mathematical measure of the chiral properties of the flux ropes or sheared-arcade structures. Over the course of evolution, reconnection as well as continuous shearing and twisting of magnetic field in such structures can drive them toward instability \citep{2008A&A...480..255R,2010SSRv..151..333M}. Upon successful eruption, newly reconnected field moves outward and is ejected to the outer corona \citep{2006ApJ...641..590G}.

Observationally, pre-eruption flux ropes are linked with solar filaments or prominences -- one of the most common large-scale magnetic features observed in the solar corona \citep{2014LRSP...11....1P}. Within a filament, dense chromospheric plasma, which is much cooler than the $1$\,MK hot corona, stays confined by the helical magnetic-field lines of the associated flux rope. Thus they appear as dark elongated structures against the bright solar disk when observed in the H$\alpha$ absorption line. The same structures look brighter on the solar limb in the H$\alpha$ emission line and are known as prominences. Instabilities in the flux rope trigger eruption of the filament, causing the confined plasma to be released in an explosion resulting in a coronal mass ejection (CME) \citep{2004ApJ...614.1054J,2013AdSpR..51.1967S}. Highly energetic events such as CMEs can have a significant impact on space weather by changing its radiative, electromagnetic, and particulate environment drastically \citep{2015AdSpR..55.2745S}. Thus, with the increasing importance of predicting space weather, understanding the evolution of magnetic-flux ropes is increasingly relevant.  

Since the coronal magnetic field is strongly affected by the distribution and complexities of the surface magnetic field, the number of filaments forming in the corona, in general, follows the sunspot cycle \citep{2015ApJS..221...33H}. However, during solar minimum, when there is little new sunspot emergence, a significant number of filaments (as high as $200$ in a year: \citeauthor{2015ApJS..221...33H}, \citeyear{2015ApJS..221...33H}) can be observed at higher latitudes ($> 30 ^{\circ}$). These are known as polar-crown filaments \citep{2014LRSP...11....1P}, and they comprise quiescent filaments forming over long neutral lines passing across the diffused and weak magnetic-field distribution. The eruption of such filaments is known to result in the occasional CMEs recorded during solar minimum \citep{2012LRSP....9....3W}. 

Using a magnetofrictional approach, this current study focuses on how eruptions are generated due to the gradual injection of non-potentiality through surface motion and magnetic reconnection in the corona during a very low activity period of Sunspot Cycle 24. In the following, we first provide a brief description of the computational model, the details of the period of our study, and different analysis tools utilised in this work in Sections \ref{sec2}. Section \ref{sec3} comprises the results obtained, which we further divide into several subsections to address different aspects of our findings. Finally, we summarise and interpret our results in Section \ref{sec4}.     

\section{Computational Model and Analysis Tools}
\label{sec2}
\subsection{Coronal Magnetic-Field Model}
\label{sec2.1}
The computational model used in this study is a combination of a surface-flux transport model and a non-potential coronal model, where the magnetic field [$\vec{B}$] within the corona evolves in response to the large-scale shearing velocity on the solar surface. The approach was introduced by \cite{2000ApJ...539..983V}, and extended to the global corona by \cite{2008ApJ...680L.165Y}. For the coronal part, we employ 
a magnetofrictional modelling approach and solve the non-ideal form of the induction equation in terms of magnetic vector potential [$\vec{B} = \curl \vec{A}$], 

\begin{equation}
\frac{\partial \vec{A}}{\partial t} = -\vec{E} 
\label{eq1}
\end{equation}

\noindent where, $\vec{E} =  - \vec{v} \times \vec{B} + \vec{N}$. Here, $\vec{E}$ represents the electric field, and $\vec{N}$ corresponds to the non-ideal part of Ohm's law. This non-potential coronal model is a simplified version of full-scale magnetohydrodynamic (MHD) models. We retain the induction Equation [\ref{eq1}] but use a ``frictional'' velocity rather than coupling it to the full momentum equation. The computational domain is within $\rm{R_{\odot} \le r \le 2.5\,R_{\odot}}$ and the full extent of co-latitudes and longitudes varying from 0\,--\,180 and 0\,--\,360 degrees respectively. We solve Equation [\ref{eq1}] using a finite-difference method on an equally-spaced grid of $360 \times 180 \times 60$ cells in longitude, sine(co-latitude), and $\log(r/\rm{R_{\odot}})$. 

The frictional velocity [$\vec{v}$] is proportional to the Lorentz force that drives the magnetic field to relax toward a force-free equilibrium, [$\vec{j} \times \vec{B} = 0$], with $\vec{j}$ and $\vec{B}$ being the current density and the magnetic field. Accordingly, the velocity field within the corona is modelled through the equation

\begin{equation}
    \vec{v} = \frac{\vec{j} \times \vec{B}}{\nu |\vec{B}|^2} + v_{\mathrm{out}}(r)\hat{\vec{e_r}}. 
    \label{eq2}
\end{equation}

\noindent In the above equation, $\nu$ is the friction coefficient and has the functional form, $\nu = \nu_0 r^2 \sin^2 \theta$ ($\theta$: co-latitude), with $\nu_0 = 3.6 \times 10^{-6}$\,s$^{-1}$. The spatial dependency of $\nu$ facilitates reduced computational time. However, at the inner boundary, the frictional velocity is set to be zero. The second term in Equation [\ref{eq2}], $v_{\rm out}(r) = v_0(r/\rm{R_\odot})^{15}$ mimics the presence of solar wind in the corona and ensures that magnetic-field lines become radial beyond $2.5\,\rm{R_{\odot}}$. For most of our simulations, we consider a solar wind with the maximum speed $v_0 = 100$\,km\,s$^{-1}$. 

At the inner boundary, however, the velocity field is replaced by two large-scale plasma flows, differential rotation and meridional circulation, along with a supergranular diffusion term modelling the net effect of unresolved small-scale convection. These flows are considered to be time-independent. The time-averaged observed values on the photosphere determine their spatial profiles and amplitudes. These flows are part of the surface-flux-transport model functioning at the inner boundary ($r = \rm{R_{\odot}}$), which controls the evolution of the radial component of magnetic field at $r = \rm{R_{\odot}}$ (for more details, see Section 2.2 of \citeauthor{2014SoPh..289..631Y}, \citeyear{2014SoPh..289..631Y}). The parameters involved in the surface-flux-transport model are chosen according to the standard values suggested by \cite{2017A&A...607A..76W}. 

We try two different forms of the non-ideal term [$\vec{N}$] in Ohm's law: ohmic diffusion or fourth-order hyperdiffusion. In the case of ohmic diffusion,

\begin{equation}
    \vec{N} = \eta_0 \left( 1 + c \frac{|\vec{j}|}{\mathrm{max}|\vec{B}|}\right) \vec{j}
    \label{eq3}
\end{equation}

\noindent while $c > 0$ ensures enhanced diffusivity in strong current sheets. The constant $\eta_0$ (we choose a value of $6 \times 10^{11}$\,cm$^2$\,s$^{-1}$) determines the effectively of the ohmic diffusion. The term, $\mathrm{max}|\vec{B}|$ denotes the maximum amplitude of $|\vec{B}|$. However, recent works with coronal magnetofriction models consider unresolved small-scale fluctuations as the major contributor to $\vec{N}$ and utilise hyperdiffusion such that

\begin{equation}
    \vec{N} = \frac{- \vec{B}}{|\vec{B}|^2} \nabla (\eta_{\mathrm{h}} |\vec{B}|^2 \nabla \alpha).
    \label{eq4}
\end{equation}

\noindent In the above equation, $\alpha = \vec{j}.\vec{B}/|\vec{B}|^2$, is the current-helicity density and the chosen value of $\eta_{\mathrm{h}}$ is $10^{31}$\,cm$^4$\,s$^{-1}$. This form of hyperdiffusion preserves magnetic helicity density, [$\vec{A}\cdot\vec{B}$], in the volume \citep{2008ApJ...682..644V}. It reduces gradients in $\alpha$ so that the coronal magnetic field evolves towards a linear force-free configuration. However, due to the large-scale shearing flows on the surface, that force-free state is never achieved in full-Sun simulations. For the majority of the current study, we consider hyperdiffusion to model $\vec{N}$.

\subsection{Period of Study}
\label{sec2.2}
We simulate continuously the period between 26 August 2018 and 22 February 2019 -- covering 180 days with a daily cadence to output 3D coronal magnetic field. However, we note that the dates do not hold much importance in our current study. The main objective is to analyse how non-potentiality builds up within six months through the magnetofrictional evolution of the coronal magnetic field and not to draw a comparison between the simulated and observed corona during this period. To initiate the simulation, we require three-dimensional magnetic-field information in the corona. Accordingly, we perform a potential-field source-surface (PFSS) extrapolation \citep{1969SoPh....6..442S,1969SoPh....9..131A} on the radial component of the observed surface magnetic-field distribution corresponding to Carrington rotation 2207 (which began on 6 August 2018 and ended on 2 September 2018). This was taken from the radial-component, pole-filled map derived by \cite{2011SoPh..270....9S} using the data from the Helioseismic and Magnetic Imager (\textit{HMI}) of Solar Dynamics Observatory (\textit{SDO}). This pole-filled map is corrected for the erroneous data near the poles, which, in general, appears due to projection effects. The PFSS extrapolation used our finite-difference code \citep{anthony_yeates_2018_1472183}. Additionally, we assume no new active regions emerged on the surface during these six months. Our selected period was quite close to the minimum of Sunspot Cycle 24; thus, only thirteen sunspots were discarded due to our assumption. Moreover, during this 180-day-long simulation, we choose certain epochs to perform additional simulations with hourly cadence to study the evolution of particular non-potential structures of interest in more detail.
 
\subsection{Different Measures of Non-Potentiality}
\label{sec2.3}
\subsubsection{Free Magnetic Energy and Current Density}
\label{sec2.3.1}
To understand the dynamics of the coronal magnetic field, we study the temporal evolution of various quantities, especially those reflecting the build-up of non-potentiality within the corona. One such measure is free energy, which is an upper limit for the expendable energy available for eruptions of non-potential structures such as flux ropes. It is computed as the difference between the non-potential solution and the corresponding solution from the PFSS model at the same point of time. Another important measure is the mean current density per unit volume within the corona. However, one must note that both of these measures are inadequate to represent the spatial distribution of non-potentiality in the coronal magnetic field. Free energy only makes sense as a global measure and cannot be defined locally. Although, the current density can be defined locally, it is not an ideal invariant and can be changed even if the magnetic field on the boundaries is fixed. So a more robust local measure, which is also ideal invariant, is required (see Section \ref{sec2.3.2}).

\subsubsection{Magnetic Helicity}
\label{sec2.3.2}
Another means of assessing non-potentiality within the corona is to calculate the relative magnetic helicity \citep{1984JFM...147..133B}. However, it also cannot provide any local-helicity information associated with non-potential magnetic structures. Recent studies \citep{2016A&A...594A..98Y,2017ApJ...846..106L} have demonstrated that evaluating the field-line helicity is an excellent way to identify the spatial distribution of magnetic helicity within solar corona. In particular, those magnetic-field lines that are significantly twisted and sheared always have strong field-line helicity, and such field lines are often spatially concentrated to form structures such as flux ropes. 

Field-line helicity is defined as the normalised magnetic helicity within an infinitesimally thin tube around a field line and is calculated through the line integral

\begin{equation}
    \mathcal{A} = \int_{L(x)} \frac{\vec{A}\cdot\vec{B}}{|\vec{B}|}  \,\mathrm{d}l.
\label{eq5}
\end{equation}

\noindent Here $l$ represents arc length along the field line $L(x)$ through point $x$, and $\vec{A}$ is a vector potential for the magnetic field $\vec{B}$. The quantity $\mathcal{A}$ can also be thought of as the flux linked with the field lines, where contributions come from twisting of magnetic-ﬁeld lines with height and winding around centres of strong ﬂux on the boundary. A more detailed theoretical basis of this quantity is discussed by \cite{2016A&A...594A..98Y} as well as \cite{2018JPlPh..84f7702Y}. If the field-line foot-points on the solar surface remained fixed in time, $\mathcal{A}$ would be an ideal invariant. However, helicity is continuously injected to the global corona through the surface motions; and thereby, we can expect a continuous evolution of field-line helicity and its preferential accumulation near non-potential structures such as flux ropes generated from magnetic reconnection.

Since the coronal volume is a magnetically open domain with non-zero $B_r$ on both the inner and outer boundaries, $\mathcal{A}$ depends on our choice of gauge for $\vec{A}$. In particular, we recompute $\vec{A}$ in a more appropriate (according to relative magnetic helicity) gauge than that given by the computation with Equation [\ref{eq1}]. The previous studies of \cite{2016A&A...594A..98Y} and \cite{2017ApJ...846..106L} computed $\mathcal{A}$ in the DeVore\,--\,Coulomb gauge $\vec{A}^{\rm DC}$, which has $A^{\rm DC}_r=0$ throughout the domain and $\nabla_h\cdot\vec{A}^{\rm DC}=0$ on the inner boundary $r=\rm{R_\odot}$. Here we use an alternative gauge $\vec{A}^*=\vec{A}^{\rm DC} + \nabla\chi$ that satisfies $\nabla_h\cdot\vec{A}^*=0$ on both the inner and outer boundaries. On closed field lines (connecting to the Sun at both ends), this gives the same  $\mathcal{A}$-values as $\vec{A}^{\rm DC}$, but on open-field lines the $\mathcal{A}$-values can differ slightly. This difference is small because the open-field lines do not tend to store field-line helicity, but it makes the calculations consistent with the poloidal-toroidal gauge of \cite{2018JPhA...51W5501B} as well as the minimal gauge condition of \cite{2018JPlPh..84f7702Y}. Moreover, unlike $\vec{A}^{\rm DC}$, this modified gauge has the property that integrating $\mathcal{A}|\vec{B}|$ over all field lines gives the relative magnetic helicity. Using this formulation, we calculate the field-line helicity for a set of coronal magnetic-field lines (with footpoints equally distributed on the photospheric boundary) based on the simulation-generated daily data. We consecutively utilise it to identify magnetic structures with high non-potentiality by applying a threshold to the field-line helicity (details are provided in Section \ref{sec3.1.2}). Moreover, the variation in the mean field-line helicity will reflect the temporal evolution of magnetic helicity in the global corona in our simulation.

Although helicity is continuously being injected into the coronal magnetic field through the shearing motion on the surface, any depletion of the mean unsigned field-line helicity in the corona is related to two factors. To understand this, consider the evolution equation for the relative helicity [$H$], which is the signed integral of field-line helicity. This may be written (cf. \citeauthor{2016A&A...594A..98Y}, \citeyear{2016A&A...594A..98Y}) as

\begin{equation}
    \frac{\rm{d}H}{\rm{d}t} = -2 \int_V \vec{N}\cdot\vec{B} \,\rm{d}V + \oint_S \vec{A} \times \big[ 2 \vec{E} + \frac{\partial \vec{A}}{\partial t}\big] \,\rm{d}\vec{S}.
    \label{eq6}
\end{equation}

\noindent First, there is an overall volume dissipation of helicity, represented by the first term on the right-hand side of Equation [\ref{eq6}]. Second, there are sudden decreases due to the ejection of unstable non-potential structures with high helicity content through the outer boundary, which we measure by integrating the second term on the right-hand side of Equation [\ref{eq6}] over a closed surface at $r = 2.5\,\rm{R_{\odot}}$. The temporal evolution of these two quantities in our simulation is discussed in Section \ref{sec3}.

\subsubsection{Change in Magnetic-Field Distribution}
\label{sec2.3.3}
The destabilisation of a non-potential structure can lead to an eruption, which is always associated with significant changes in coronal magnetic-field structure. Therefore, it should be reflected in the distribution of open magnetic field passing through $2.5\,\rm{R_{\odot}}$ as well as in the horizontal component of magnetic-field associated with the non-potential structure itself. Open magnetic-field lines are those with one end at the photospheric boundary and the other at the $2.5\,\rm{R_{\odot}}$ boundary, and the extended regions with such open-field lines are regarded as coronal holes. In fact, one of the significant precursors of the onset of a CME is dimming in coronal emission in regions surrounding the unstable flux rope associated with the CME. Several computational and observational studies have suggested that coronal dimming is linked with the formation of transient coronal holes around the non-potential structure \citep{1998ApJ...498L.179G,2000JGR...10518187G,2001JGR...10629239K,2006SoPh..238..117A,2011LRSP....8....1C}. Thus, we inspect how the coronal-hole area changes in the vicinity of the evolving non-potential structures in our simulation. These localised changes in the coronal-hole area also cause significant amplitude variation of open magnetic flux over the period of the simulation.   

At the outer boundary ($r = 2.5\,\rm{R_{\odot}}$), the magnetic field is primarily radial because of the solar wind. However, as an erupting non-potential structure migrates upwards and eventually passes through the outer boundary, the horizontal component of magnetic field [$B_{\perp} = (B_{\theta}^2 + B_{\phi}^2)^{1/2}$] at $r = 2.5\,\rm{R_{\odot}}$ becomes significantly enhanced \citep{2016A&A...594A..98Y,2017ApJ...846..106L}. Thus we also compute the variation of $B_{\perp}$ at $2.5\,\rm{R_{\odot}}$ in simulation-generated data.

\subsubsection{Magnetic Pressure and Tension Forces}
\label{sec2.3.4}
We inspect the evolution of radial magnetic-tension force and pressure gradient across individual non-potential structures of interest. Within the twisted and helical structure of a flux rope, magnetic pressure acts outward from the flux-rope core axis, whereas the tension force acts inward (see Figure 4 of \citeauthor{2014SoPh..289..631Y}, \citeyear{2014SoPh..289..631Y}). Previous magnetofriction simulations \citep{2009ApJ...699.1024Y,2014SoPh..289..631Y} have utilised this concept to identify flux ropes. Similar to \cite{2014SoPh..289..631Y}, at each point on the computational grid, we calculate the normalised radial magnetic-tension force and magnetic-pressure gradient determined by the following equations (in our simulation $\mu_0$ is equal to one),

\begin{equation}
    T_r = \frac{\rm{R_{\odot}}}{B^2} \left(\vec{B}\cdot \nabla B_{r} - \frac{B_{\theta}^2}{r} - \frac{B_{\phi}^2}{r} \right), \hspace{1cm} P_r = - \frac{\rm{R_{\odot}}}{B^2} \frac{\partial}{\partial r} \left(\frac{B^2}{2}\right).
    \label{eq7}
\end{equation}

\noindent Then to locate flux-rope structures we identify those grid points ($r_i,\theta_j,\phi_k$) where the following four conditions are satisfied:

\begin{align}
    P_r (r_{i-1}, \theta_j , \phi_k)& < - 0.4 \nonumber \\
    P_r (r_{i+1}, \theta_j , \phi_k)& > 0.4  \nonumber \\
    T_r (r_{i-1}, \theta_j , \phi_k)& >  0.4 \nonumber \\
    T_r (r_{i+1}, \theta_j , \phi_k)& < - 0.4 . 
    \label{eq8}
\end{align}

This analysis adds to our understanding of the distinctive evolving nature of different non-potential structures, which are primarily selected based on their field-line helicity content and enhancement in the $B_{\perp}$ at the outer boundary. Moreover, we particularly focus on detecting positive tension force at 2.0\,$\rm{R_{\odot}}$, which is a signature of helical magnetic-field lines associated with an unstable flux-rope-like structure passing through the coronal height of 2.0\,$\rm{R_{\odot}}$ (see Sections \ref{sec3.1} and \ref{sec3.2} for more details).

\subsubsection{Quantitative Analysis of Individual Non-Potential Structures}
\label{sec2.3.5}
We intend to quantify different physical properties, such as total area, total magnetic flux, and total field-line helicity content associated with individual non-potential structures formed during our coronal magnetic-field simulation. As these structures sometimes undergo drastic reformation or disappear entirely due to instability, we are also interested in corresponding changes in these physical quantities during those epochs. An analysis of all of the erupting structures in our simulation is provided in Section \ref{sec3.3}. The structures are selected based on a field-line helicity threshold, i.e. determined from the magnetic field. They are not selected specifically by physical size, which in any case might appear different in observations where structures would be determined primarily by plasma density rather than magnetic field.

\subsubsection{Dependency on Model Parameters}
\label{sec2.3.6}
To test the robustness of our results, we perform additional magnetofriction simulations of 180 days, starting with the same initial surface magnetic-field map. However this time, we consider ohmic diffusion to model $\vec{N}$ (using Equation [\ref{eq3}]). We again evaluate different measures of non-potentiality over 180 days, similar to our analyses with hyperdiffusion. Another critical factor that may influence our results is the speed of the solar wind used in the magnetofriction simulations ($v_{\mathrm{out}}$ in Equation [\ref{eq2}]). Thus we execute additional simulations of 180 days with a slower solar wind (maximum speed 50\,km\,s$^{-1}$, and $\vec{N}$ modelled according to hyperdiffusion). The corresponding results are briefly discussed in Section \ref{sec3.4} with further details in the Appendix.

\begin{figure}[!ht]    
\centerline{\includegraphics[width=0.7\textwidth,clip=]{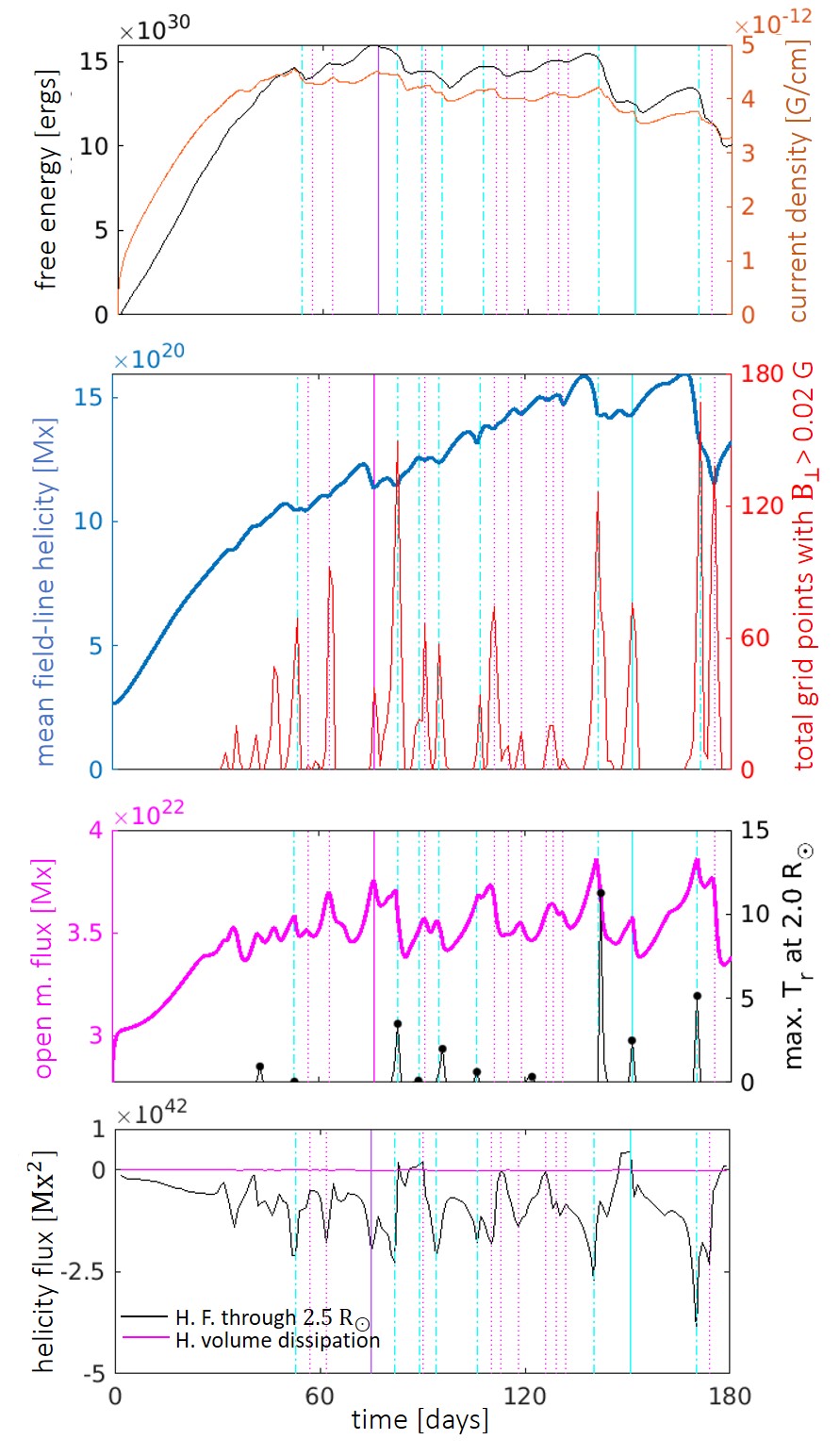}}
\caption{Global measures. Temporal evolution of free energy and current density are depicted by the black and red curves, respectively, in the top row. The second row shows the temporal evolution of mean field-line helicity, over-plotted with the total number of grid points on the outer boundary where $B_{\perp} > 0.02$\,G. The third row shows the evolution of open magnetic flux and the maximum amplitude of positive radial-tension force. In the last row, the helicity volume dissipation and helicity flux through the outer coronal boundary, from Equation [\ref{eq6}], are presented. The vertical lines correspond to the 19 identified epochs of sudden but significant changes of non-potential measures associated with eruptive ``events'' of different classes: flux-rope eruptions (cyan lines) and overlying-arcade eruptions (magenta lines). Two of them (marked by the solid cyan and magenta-vertical lines) are discussed in more detail.}
\label{fig1}
\end{figure} 
 
\section{Results}
\label{sec3}
\subsection{Generation and Evolution of Non-Potentiality}
\label{sec3.1}
\subsubsection{Global Measures}
\label{sec3.1.1}
First, we focus on the temporal evolution of global measures such as the free magnetic energy and current associated with the build-up of non-potentiality within the corona. The first row in Figure \ref{fig1} depicts the variation of free magnetic energy and volume-averaged unsigned current density in the corona. Both quantities are strongly correlated and have a general decreasing trend beyond the initial rising phase. This initial phase arises because the initial potential coronal magnetic field requires a certain amount of ``ramp time'' to form non-potential structures. In our magnetofriction simulation, this ramp time is about 50 days. After this, and since our simulation does not include emergence of new sunspots, the total photospheric magnetic-field strength decreases monotonically over time, thereby resulting in the observed decay of free magnetic energy and current. However, several sharp decrements at certain epochs are noteworthy, which we inspect thoroughly in the following section.

We calculate the field-line helicity for a set of magnetic-field lines in the corona from the daily simulated coronal magnetic field according to the method described in Section \ref{sec2.3.2}. The second row in Figure \ref{fig1} represents the evolution of the average (unsigned) field-line helicity in the global corona, which again shows sharp changes similar to the current and free-energy evolution at the same epochs. On each day of the 180-day simulation, from the $B_{\perp}$ maps generated on the outer boundary ($r = 2.5\,\rm{R_{\odot}}$), we record the number of grid points with $B_{\perp} > 0.02$\,G, the temporal evolution of which is depicted by the red curve on the second row of Figure \ref{fig1}. This shows strong peaks around the time when the mean helicity drops. There are also peaks at similar times in the open magnetic flux (see the third row in Figure \ref{fig1}). 

As mentioned in Section \ref{sec2.3.4}, we calculate the radial magnetic-tension force in the coronal volume and primarily search for positive values of $T_r$ at 2.0\,$\rm{R_{\odot}}$. The reasoning is as follows: a flux-rope structure comprises helical magnetic-field lines, and the lower part (with respect to the flux-rope axis) of these field lines must contribute to positive $T_r$. It should be detected in the corona as an unstable flux rope erupts and rises. In our $T_r (\theta, \phi)$ analysis at a fixed radius, we choose the coronal height at 2.0\,$\rm{R_\odot}$, which is somewhere between the inner and outer boundaries (closer to the outer boundary). In the third row of Figure \ref{fig1}, the black curve with distinct peaks represents the maximum amplitude of positive $T_r (r=2.0\,\rm{R_{\odot}}, \theta, \phi)$. We find that for particular cases (indicated by cyan lines), the epochs according to the maximum $T_r$ match well with those obtained from other non-potential measures, such as $B_{\perp}$, mean field-line helicity, etc.

\begin{figure}[ht!]    
   \centerline{\hspace*{0.015\textwidth}
               \includegraphics[width=0.5\textwidth,clip=]{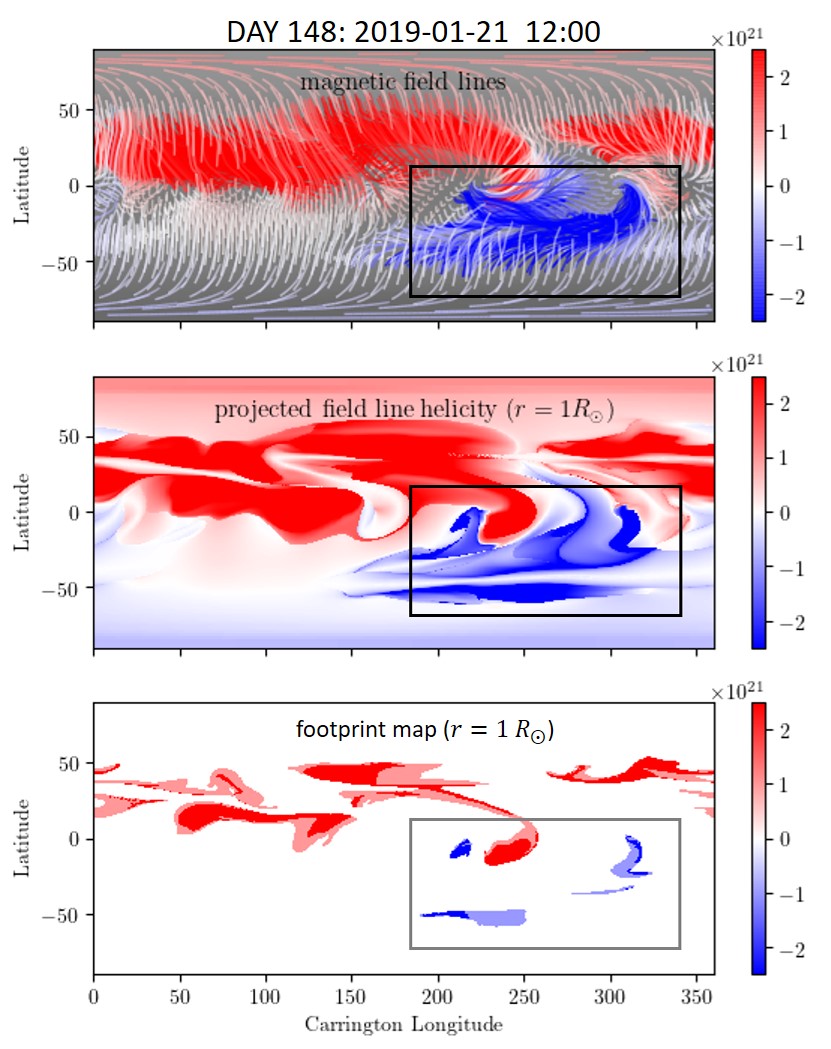}
               \hspace*{-0.03\textwidth}
               \includegraphics[width=0.5\textwidth,clip=]{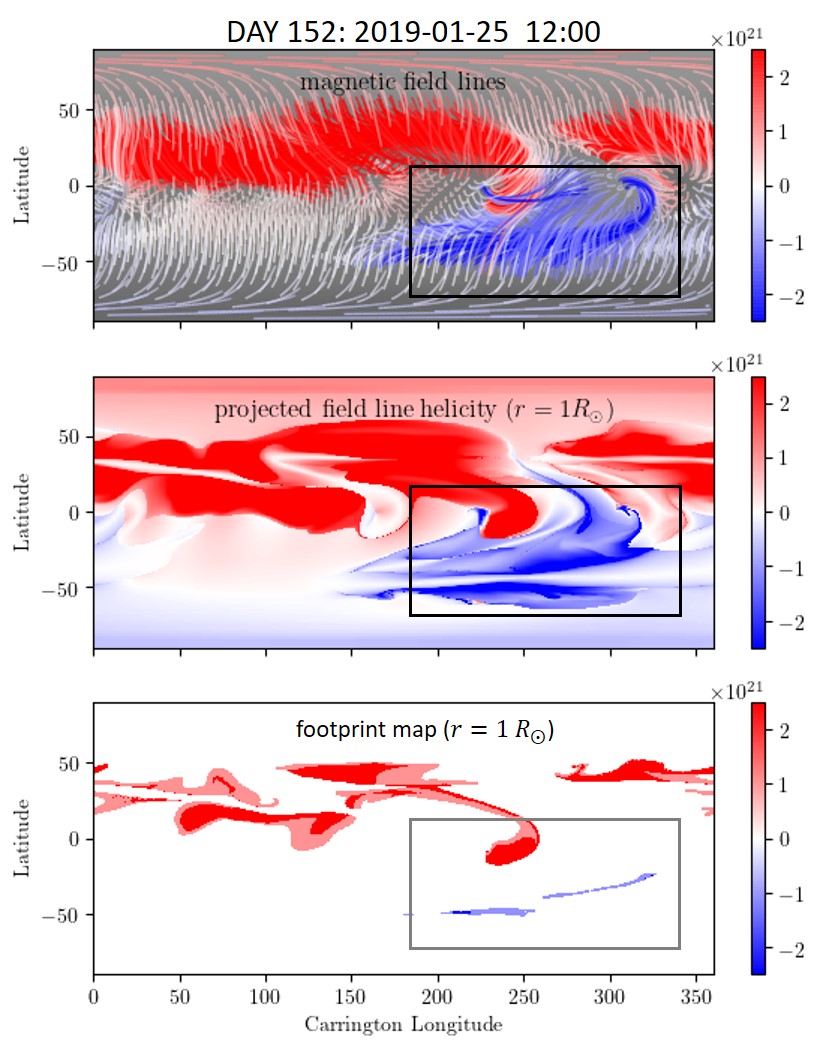}
              }
     \vspace{-0.35\textwidth}   
     \centerline{\Large \bf     
         \hfill}
     \vspace{0.31\textwidth}    
   \centerline{\hspace*{0.015\textwidth}
               \includegraphics[width=0.5\textwidth,clip=]{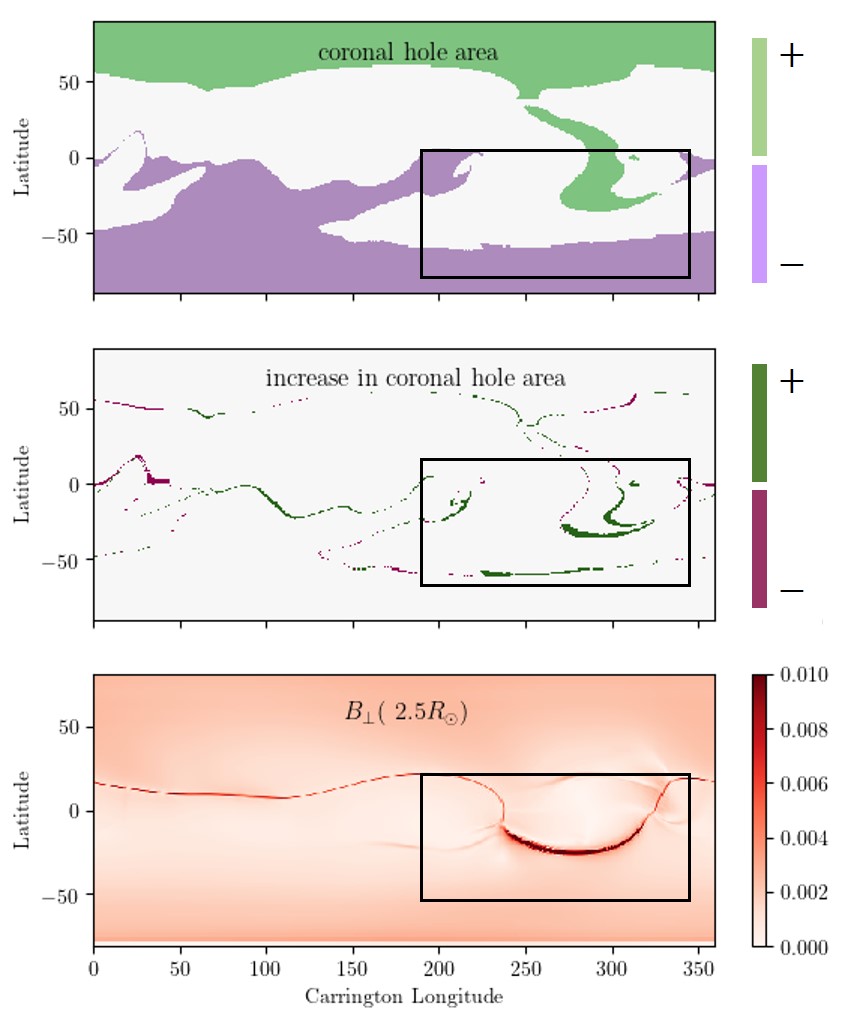}
               \hspace*{-0.03\textwidth}
               \includegraphics[width=0.5\textwidth,clip=]{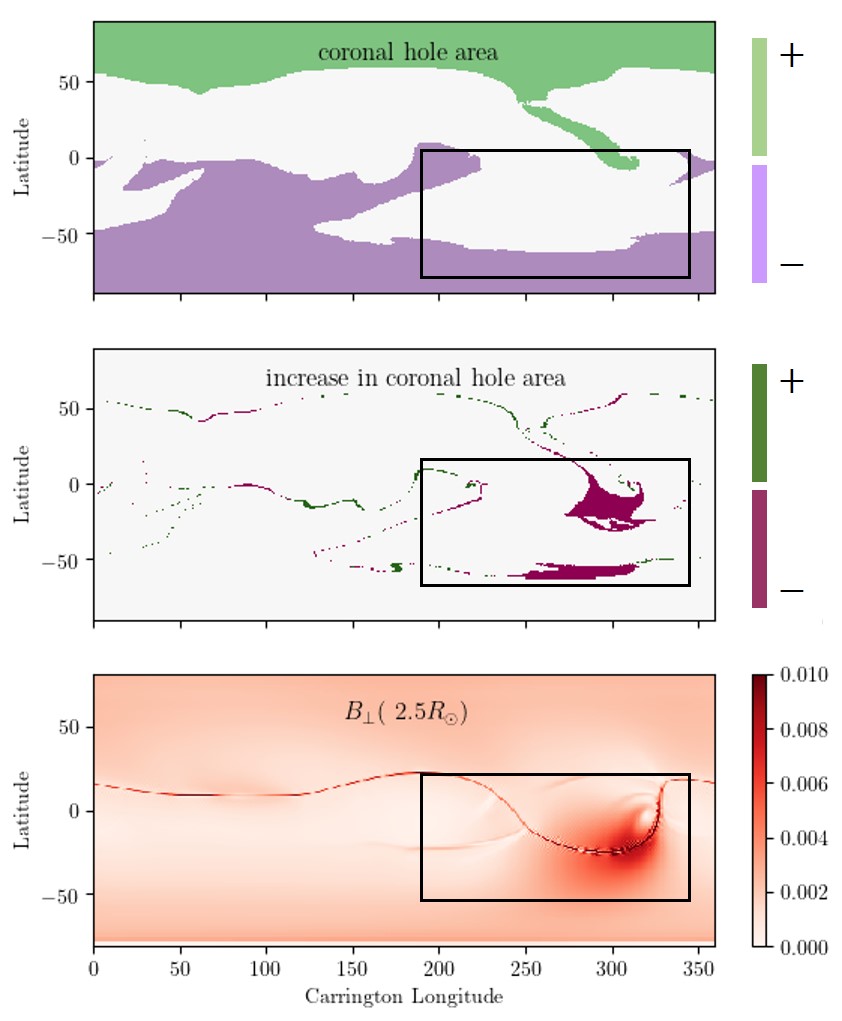}
              }
     \vspace{-0.35\textwidth}   
     \centerline{\Large \bf     
         \hfill}
     \vspace{0.32\textwidth}    
              
\caption{Spatial distributions on two nearby dates: the first row depicts a top view of the magnetic-field lines colour-coded according to their field-line helicity (positive helicity in red and negative in blue, units in Mx). The grey background corresponds to the radial component of the photospheric magnetic field (within $\pm$ 10\,G). The second row shows the photospheric mapping of field-line helicity. The third row represents the footprint of selected non-potential structures, marking cores and their extension with darker and lighter shades, respectively. The foot-points of the open-field lines with upward (green) and downward (violet) directions are shown on the fourth row. The fifth row depicts the change in coronal-hole area compared to the previous day; dark green and dark purple suggest opening up and closing down of field lines, i.e. increase and decrease in coronal-hole area, respectively. The distribution of the horizontal component of the magnetic field [G] at the outer boundary is depicted in the last row.}
\label{fig2}
\end{figure}

Lastly, we study the global quantities associated with the change in magnetic helicity, as discussed in Section \ref{sec2.3.2}. The last row in Figure \ref{fig1} shows the temporal evolution of the dissipation term and the helicity flux through the outer boundary, where variation in the former is negligible compared to the latter, owing to the nearly ideal nature of the simulation. The helicity flux shows a mean background value, due to the differential rotation of open-field lines \citep{2016A&A...594A..98Y}, superimposed with significant fluctuations. These are associated with those that have already been mentioned in other global quantities.

\subsubsection{Local Measures: Spatial Distribution of Non-Potentiality}
\label{sec3.1.2}

Among the different global measures of non-potentiality, magnetic-field-line helicity, coronal-hole area, and $B_{\perp}$ maps generated on the outer boundary ($r = 2.5\,\rm{R_{\odot}}$) can reveal how the non-potential magnetic field is spatially distributed within the corona. Thus we inspect the spatial and temporal evolution of these quantities during our 180-day simulation with daily cadence. Note that radial magnetic pressure gradient and tension force maps can also provide similar information, which we have utilised later in Section \ref{sec3.2}. 

The first row of Figure \ref{fig2} shows the two-dimensional projection of magnetic-field lines on the solar surface with colours (red-white-blue) assigned according to their field-line helicity. This figure's left and right columns correspond to the corona on 21 and 25 January 2019 (Days 148 and 152), respectively. We further map the field-line helicity at the field-line footpoints, using an equally spaced grid in sine-latitude and longitude on the photospheric boundary following the same technique utilised by \cite{2017ApJ...846..106L}. The results are shown in the second row of Figure \ref{fig2}. Notably, the distribution of $\mathcal{A}$ on the surface is such that distinct domains are formed around the footpoints of more complex and twisted field structures. These maps are then further used to detect footpoints of non-potential structures through implementing a threshold technique based on the intensity of field-line helicity. Following the same method described in \cite{2017ApJ...846..106L}, we apply two thresholds to identify both the strong core and outer extension of the structure, [$\tau_{\rm{c}}$] and [$\tau_{\rm{e}}$] respectively. The amplitude of $\tau_{\rm{c}}$ and $\tau_{\rm{e}}$ are chosen according to the following relations,

\begin{equation}
\tau_{\mathrm{c}} = \frac{\overline{\mathcal{A}(t)}}{\overline{\mathcal{A}}_\mathrm{ref}} \tau_{\mathrm{c,ref}}; \hspace{0.1cm} \hspace{0.1cm} \rm and \hspace{0.1cm} \tau_{\mathrm{e}} = \frac{\overline{\mathcal{A}(t)}}{\overline{\mathcal{A}}_\mathrm{ref}} \tau_{\mathrm{e,ref}}
\label{eq9}
\end{equation}

\noindent where $\overline{\mathcal{A}(t)}$ is the mean unsigned field-line helicity, $\overline{\mathcal{A}}_\mathrm{ref} = 1.29 \times 10^{21}$\,Mx, $\tau_{\mathrm{c,ref}} = 4.84 \times 10^{21}$\,Mx, and $\tau_{\mathrm{e,ref}} = 3.39 \times 10^{21}$\,Mx. These particular values were suggested by \cite{2017ApJ...846..106L} based on their careful and thorough calibration to detect flux ropes from a similar magnetofriction simulation covering a period of eighteen years. The mean magnetic flux of erupting structures in their simulation was about $10^{21}$\,Mx, which is close to that of the observational estimates, between $10^{19}-10^{22}$\,Mx \citep{2005JGRA..110.8107L}. These estimates can be compared with those estimated from observations of interplanetary magnetic clouds. Although helicity for magnetic clouds is not very well constrained due to the uncertainty in estimating the associated length of the flux rope in the heliosphere, \cite{2017ApJ...846..106L} found that the mean helicity of a typical erupting structure in their simulation was reasonable and consistent with magnetic-cloud observations (more details are provided in Section \ref{sec3.3}). The third row in Figure \ref{fig2} depicts detected structures where the helicity magnitude exceeds these thresholds in the projected-helicity maps on 21 and 25 January 2019. These identified non-potential structures selected based on the field-line-helicity threshold are likely to have twisted magnetic-field lines similar to flux ropes (for an example see our first case study in Section \ref{sec3.2.1} below). However, we have found some instances where the detected structure has a strongly sheared arcade rather than helically twisted field lines -- we will see an example in our second case study below (Section \ref{sec3.2.2}).

Note that the sign of helicity is primarily positive in the northern and negative in the southern hemisphere, which is precisely opposite to the predominant hemispheric pattern of magnetic helicity observed in solar filaments. This disparity in our simulation is caused by the differential rotation being the sole process to determine the nature of helicity \citep{1997SoPh..175...27Z}. \cite{2012ApJ...753L..34Y} have shown that in the absence of helicity transport from emerging active regions, high-latitude flux ropes are bound to have oppositely signed helicity, especially in the declining phase of a solar cycle. Thus the opposite helicity pattern is quite expected from our simulation, which excluded emergence of new sunspots and was performed for a period very close to the cycle minimum. Nevertheless, our simulation clearly predicts the growth of non-potentiality in the coronal magnetic field and the eventual formation of flux ropes through reconnection. Thus we expect our analyses and findings will still be relevant in case of a longer simulation with the inclusion of active regions, but likely with the opposite sign of helicity at some locations.  

A closer look at the first three rows of Figure \ref{fig2} reveals a noteworthy change of coronal structures from 21 to 25 January 2019, highlighted by a rectangular box in each of the figures. Primarily, on the third row, we notice the disappearance of two blue structures near $200$ and $320$ degrees longitude in the southern hemisphere (initially extended within zero to $- 50$ degrees latitude on 21 January 2019). Simultaneously, field line distribution (see the first row) changed significantly around the same locations, which correspond to the two footpoints associated with a coronal non-potential structure (see the set of blue field lines disappearing in the top row of Figure \ref{fig2}). Furthermore, the disappearance of that particular structure is also reflected in the temporal evolution of the mean field-line helicity (see the second row of Figure \ref{fig1}). On Day 151 (which is 24 January 2019), we observe a sudden decrease in the blue curve (indicated by a cyan-vertical line). This indicates that the recurrent decreases in mean field-line helicity are likely to be associated with the disappearance and eventual eruption of non-potential structures.

As discussed in Section \ref{sec2.3.3}, an eruption causes a significant change in coronal magnetic-field structures; therefore, it should be reflected in the distribution of open magnetic field too. On the fourth row of Figure \ref{fig2}, we present coronal-hole maps of 21 and 25 January 2019 (Days 148 and 152) based on the distribution of open-field lines. A notable change is visible near $300$ degrees longitude in the southern hemisphere, almost at the same location where the non-potential structure has disappeared. We further investigate the difference in coronal-hole area between two consecutive days. The dark green patch at about $300$ degrees longitude and $-50$ degrees latitude in the first figure on the fifth row indicates the opening up of new field lines. As an eruptive non-potential structure becomes unstable and rises through the corona, the overlying field lines open up. Therefore, the coronal-hole area in the vicinity of the structure grows. This causes an increase in the overall amplitude of open magnetic flux, and indeed we notice a peak on Day 151 (see the third row of Figure \ref{fig1} indicated by solid-cyan line). In the post-eruption configuration, field lines close down, ensuing a significant decrease in coronal-hole area, consistent with the purple patch in the second image on the fifth row of Figure \ref{fig2}. Such transient changes in coronal-hole area are also reported by observational studies on the evolution of coronal dimming associated with erupting CMEs \citep{2007ApJ...660.1653M,2017MNRAS.471.4776G}. 

Additionally, we observe the strongest values of $B_{\perp}$ along the heliospheric current sheet between $240$ and $330$ degrees longitude on the left-side figure on the last row of Figure \ref{fig2} corresponding to 21 January 2019 (Day 148), which is associated with the outward moving non-potential structure. By 25 January 2019 (Day 152), the strength of $B_{\perp}$ along the heliospheric current sheet decreased, and entirely disappeared on the following Day 153. In agreement, the global evolution of $B_{\perp}$ (see the second row of Figure \ref{fig1}) shows a peak on Day 151 (24 January 2019) coinciding with the timing of probable eruption of the non-potential structure under consideration.

Lastly, in the evolution of helicity flux through the outer boundary (see the last row in Figure \ref{fig1}, indicated by a solid-cyan line), we notice around Day 151 a significant enhancement in the helicity flux through the outer boundary. This increment is causally connected with the disappearance of the negative-helicity flux rope between 21 (Day 148) and 25 January 2019 (Day 152), which removes negative helicity from the coronal volume (see the third row of Figure \ref{fig2}).

\begin{figure}[ht!]    
\centerline{\includegraphics[width=0.50\textwidth,clip=]{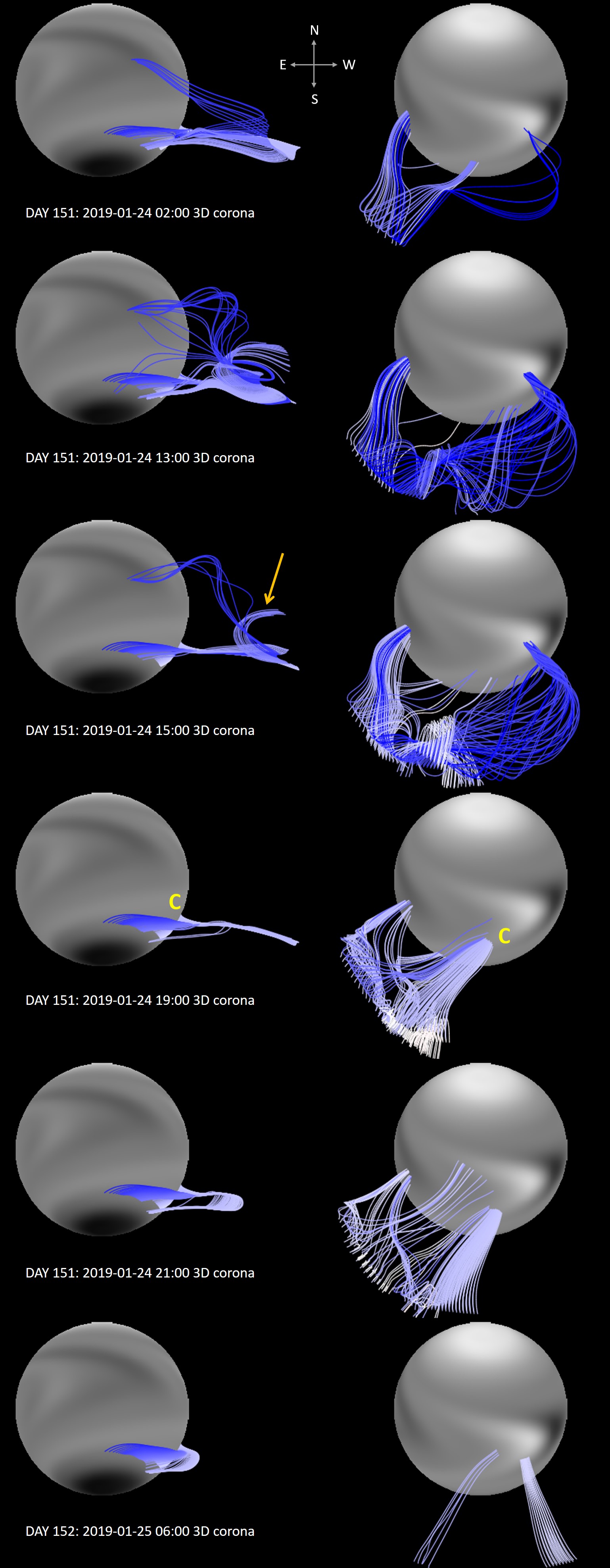}}
\caption{Evolution of the first case study (an erupting flux rope) viewed from two different viewing angles, where the field lines are colour-coded according to the field-line helicity with the maximum amplitude $2.5 \times 10^{21}$\,Mx. The colour blue represents negative field-line helicity with the darker shades corresponding to increasing amplitude. The magnetic-field distribution on the solar disk is depicted in shades of grey (within $\pm 5$\,G). Left images show field lines traced from the flux-rope footprint on the inner boundary, and right images those traced from the outer boundary.}
\label{fig3}
\end{figure}

Our analyses so far indicate that the disappearance of a non-potential structure is characterised by simultaneous model signatures, including a drop in the amplitude of mean field-line helicity, a significant change in coronal-hole area, a peak in the strength of $B_{\perp}(r = 2.5\,\rm{R_{\odot}})$ as well as in open magnetic flux, and an increase in helicity flux through the outer boundary. Globally, these changes are accompanied by decreases in free energy and current during the same epoch (as indicated in Figure \ref{fig1}). In our 180-day-long simulation, there are multiple instances where all these signatures are visible concurrently, and we label each of these instances as an individual \textit{``event''}. Taking individual peaks in $B_{\perp}$ (with threshold $ 0.02$\,G) and significant simultaneous change in field-line distribution as indicators of separate events, we identify 19 significant events, which are marked by 19 vertical lines in Figure \ref{fig1}. Note that events associated with the first four peaks in $B_\perp$ were excluded from our analyses due to their occurrence during the initial (model-induced) ramp phase. The following section describes the evolution of coronal magnetic field during each of these events to find whether all of them correspond to the eruption of non-potential structures.

\subsection{Classes of Eruption}
\label{sec3.2}
To determine the nature of the 19 events, we generate additional snapshots from the simulation with hourly cadence for a period of three to five days around the time of each event. Analysing the hourly data helps us to understand the field-line dynamics during these events in more detail and to determine their physical nature.

\subsubsection{Case Study: Flux-Rope Eruption Event}
\label{sec3.2.1}
First, we examine further the event on Day 151 that was illustrated in Section \ref{sec3.1}, as an example of a clear flux-rope eruption. In Figure \ref{fig3}, we present snapshots with three-dimensional views of the flux rope. In the figures on the left side, the plotted field-lines are selected based on the flux-rope footprint maps (shown in the third row of Figure \ref{fig2}). On the right, the figures show only those field lines passing through the outer boundary and where $B_{\perp}(r = 2.5\,\rm{R_{\odot}}) > 0.01$\,G (according to the last row of Figure \ref{fig2}). Thus increasing numbers of field lines would suggest proportional growth in the strength of $B_{\perp}$ on the outer boundary. Although the selection processes of the field lines are different for figures on the left and right columns, similarities in the footpoints of the field lines clearly demonstrate that they all belong to the same structure.

As time advances, field lines get more twisted and form the helical structure associated with a flux rope, which can be seen in the first two rows. Eventually, the structure becomes unstable, and the ejection process initiates. In the third row of Figure \ref{fig3}, we can see a few ``U''-shaped field lines with low helicity (indicated by a yellow arrow). The creation of such U-loops occurs through reconnection of field lines at the quasi-separatrix layer over the polarity-inversion line \citep{2006ApJ...642.1193M}. The U-loops are then pushed through the outer boundary by the high-speed solar wind. As the reconnection process continues, we observe strongly sheared field lines associated with the flux rope to be ejected, and new field lines with significantly less helicity to be formed below the flux rope (primarily seen in the right-side figure in the third row of Figure \ref{fig3} marked by ``C''). In the following hours, the number of field lines passing through the outer boundary decreases drastically as low-lying field lines form at lower heights (see the last two rows in Figure \ref{fig3}). Thus the disappearance of the blue patch in the third row of Figure \ref{fig2}, as well as the reduction of coronal-hole area (the fifth row in the same figure) on 25 January 2019 (Day 152), are both consistent with the hourly evolution of the non-potential structure and suggest that this event represents the full eruption of a pre-existing flux rope.

\begin{figure}[ht!]    
\centerline{\includegraphics[width=1.0\textwidth,clip=]{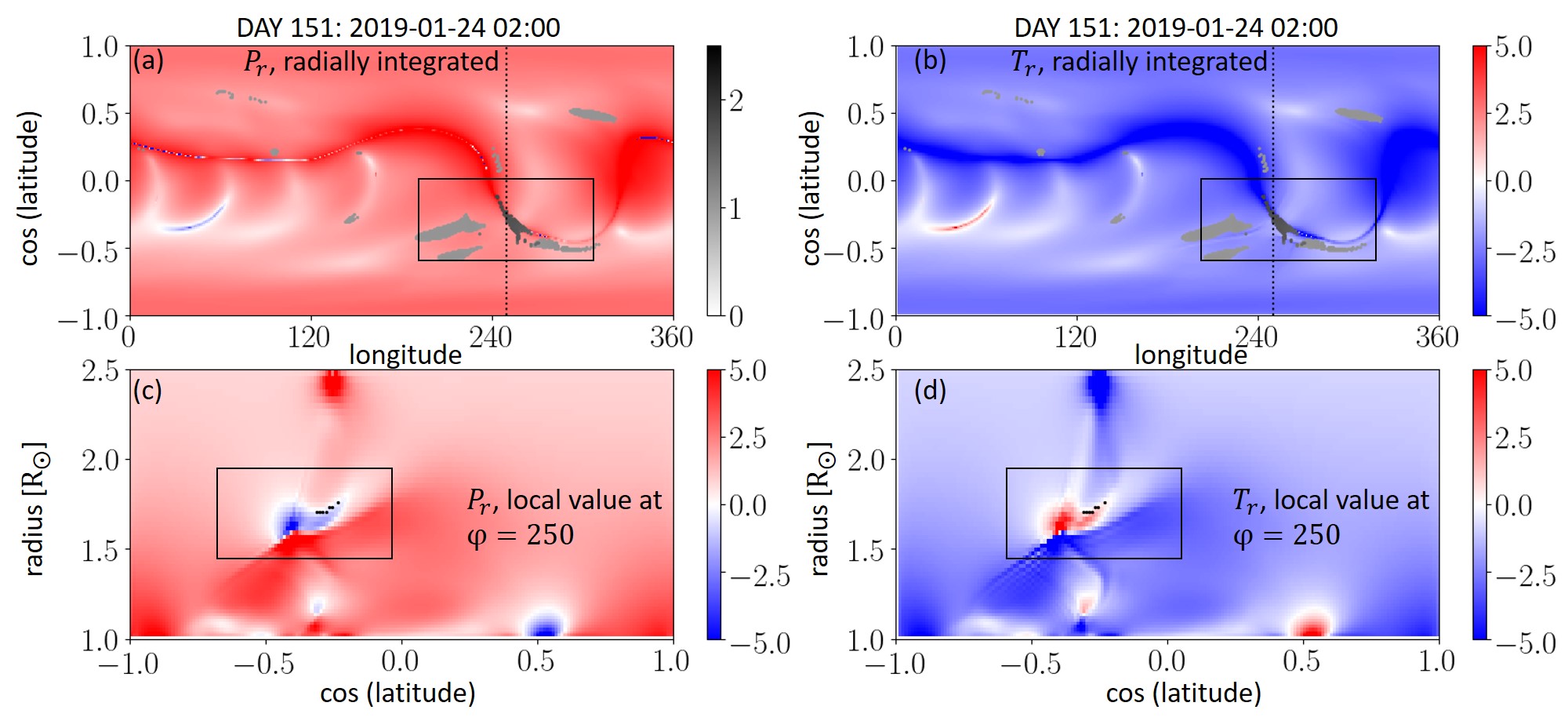}}
\caption{Radial pressure gradient and the radial component of magnetic-tension force across the flux rope in case study 1 are presented. The forces are shown by the red/blue colour map, while grid points satisfying the conditions in Equation [\ref{eq8}] are shown by dots. In (a) and (b), these are coloured by radius according to the black/white colour scale while the viewing angle is perpendicular to the surface. In (c) and (d), the forces are plotted as functions of radius and cos(latitude) across the dotted vertical cut at $250^{\circ}$ longitude.}
\label{fig4}
\end{figure}

We also analyse the evolution of radial magnetic-tension force and pressure gradient across the flux-rope structure based on the method described in Section \ref{sec2.3.4}. The first and second columns in Figure \ref{fig4} show the initial distributions of the pressure gradient and radial magnetic-tension force. On the first row, forces are radially integrated, as if looking down from the outer boundary. The colours of the identified flux-rope points in Figure \ref{fig4} [a] and [b] indicate their height above the photosphere. On the bottom row, the figures depict the value in a single meridional cut, which was chosen at longitude $250^{\circ}$. The black dots on the figures on these rows show the identified flux-rope points at that particular longitude. 

\begin{figure}[ht!]    
\centerline{\includegraphics[width=1.0\textwidth,clip=]{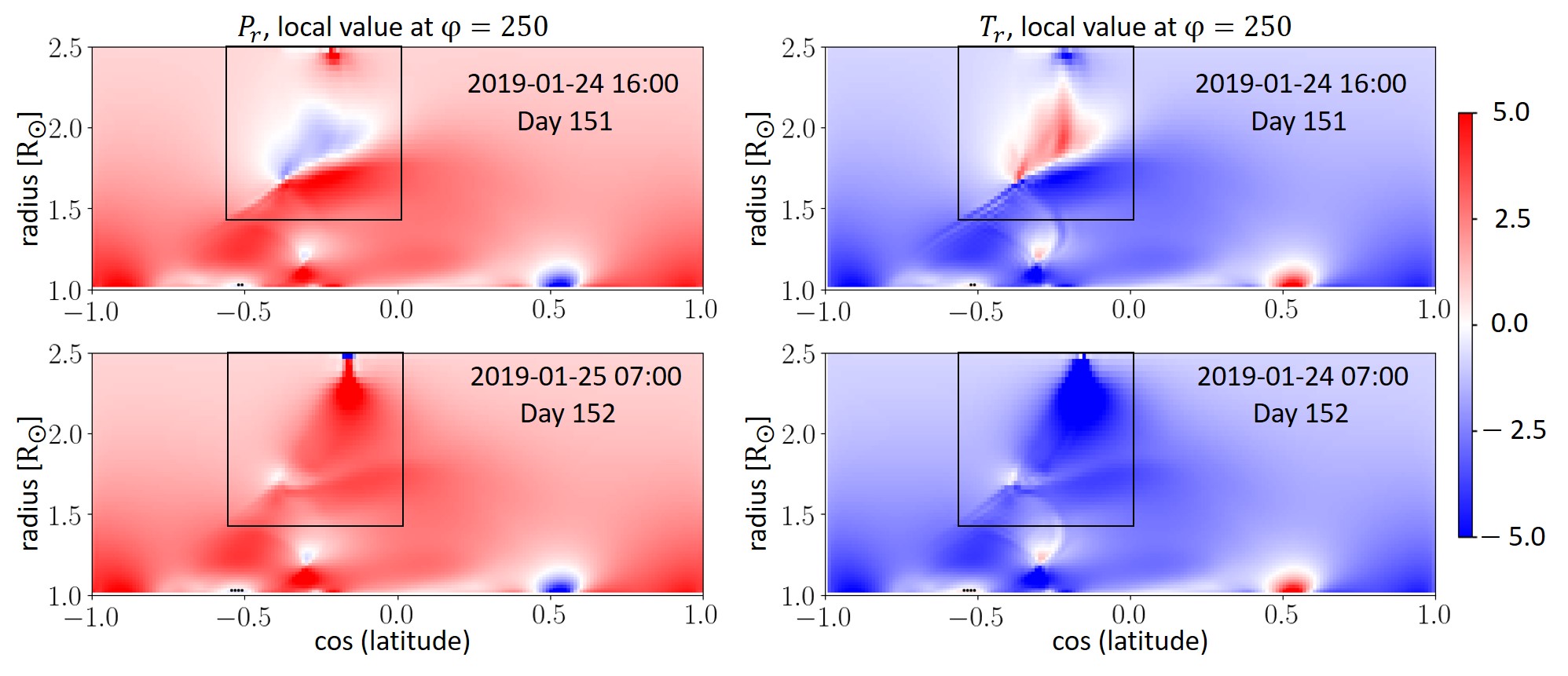}}
\caption{Next stages of evolution of pressure-gradient and radial-tension force across the flux rope (at $250^{\circ}$ longitude) with the values varying between $\pm 5.0$ for both the pressure-gradient and radial-tension force (shown by the colour bar on the right).}
\label{fig5}
\end{figure}

The initial position (as evaluated on 24 January 2019, Day 151, at 02:00) of the erupting flux rope is highlighted by rectangular boxes in each of the maps in Figure \ref{fig4}. We can clearly see that the pressure and tension forces are oppositely directed within the flux rope. In the later stage of evolution, the flux rope starts to migrate through the corona as visible in the top row of Figure \ref{fig5} and eventually is ejected through the outer boundary. On the last row of Figure \ref{fig5}, we find no trace of the flux rope, indicating a full-scale eruption, which is quite consistent with the earlier analyses of the same event. Moreover, focusing on the maps of the tension force only, we can see significant positive values of $T_r$ at the coronal height, 2.0\,$\rm{R_{\odot}}$, which corresponds to the peak on Day 151 in maximum positive $T_r (r = 2.0\,\rm{R_{\odot}}, \theta, \phi)$ (see the black curve in the third row of Figure \ref{fig1}, indicated by the solid-cyan line).

\subsubsection{Case Study: Overlying-Arcade Eruption Event}
\label{sec3.2.2}

We perform qualitative examination of the 3D magnetic-field-line distribution during the hourly evolution of each of the 19 events and search for helical structures similar to the event discussed in Section \ref{sec3.2.1}. We also check whether $T_r$ has positive values near $2.0\,\rm{R_{\odot}}$. Our analysis shows that, unlike the event discussed in Section \ref{sec3.2.1}, many of the 19 events do not involve the ejection of a helical flux rope, although each of them coincides with significant drops in the mean field-line helicity and increases in open flux and $B_{\perp}(r = 2.5\,\rm{R_{\odot}})$. Here, we present a similar analysis to Section \ref{sec3.2.1} for one such event before studying its hourly evolution.

\begin{figure}[ht!]    
   \centerline{\hspace*{0.015\textwidth}
               \includegraphics[width=0.5\textwidth,clip=]{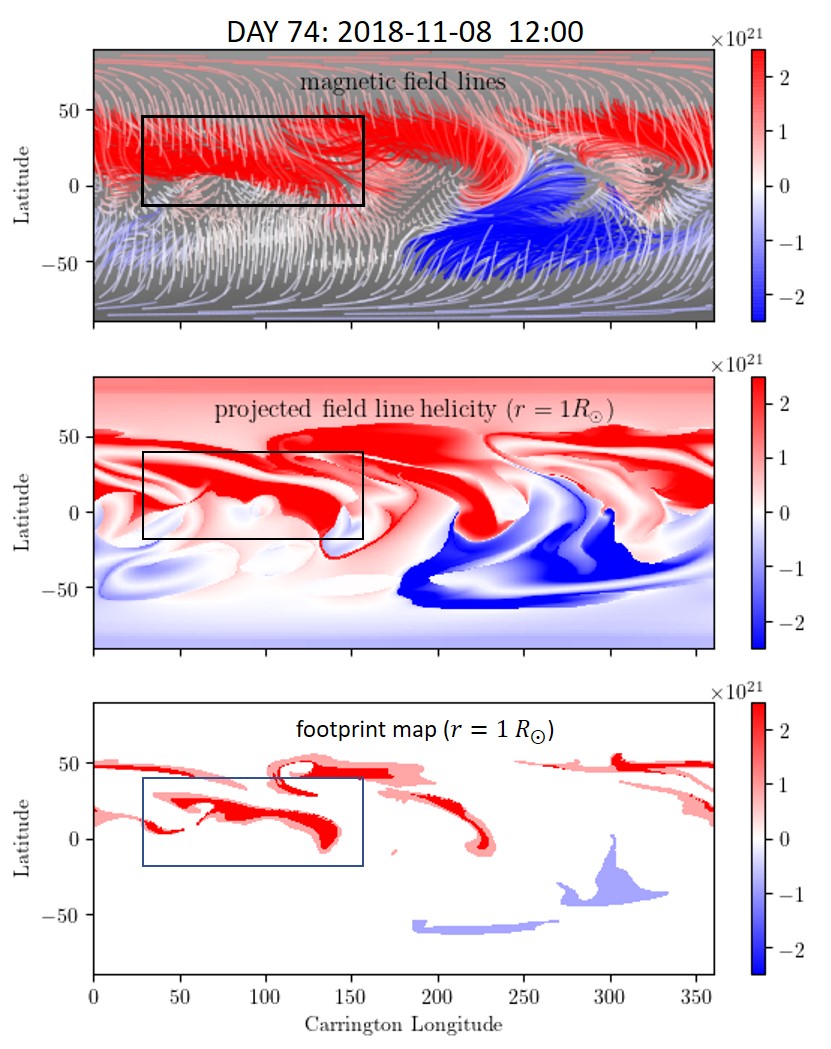}
               \hspace*{-0.03\textwidth}
               \includegraphics[width=0.5\textwidth,clip=]{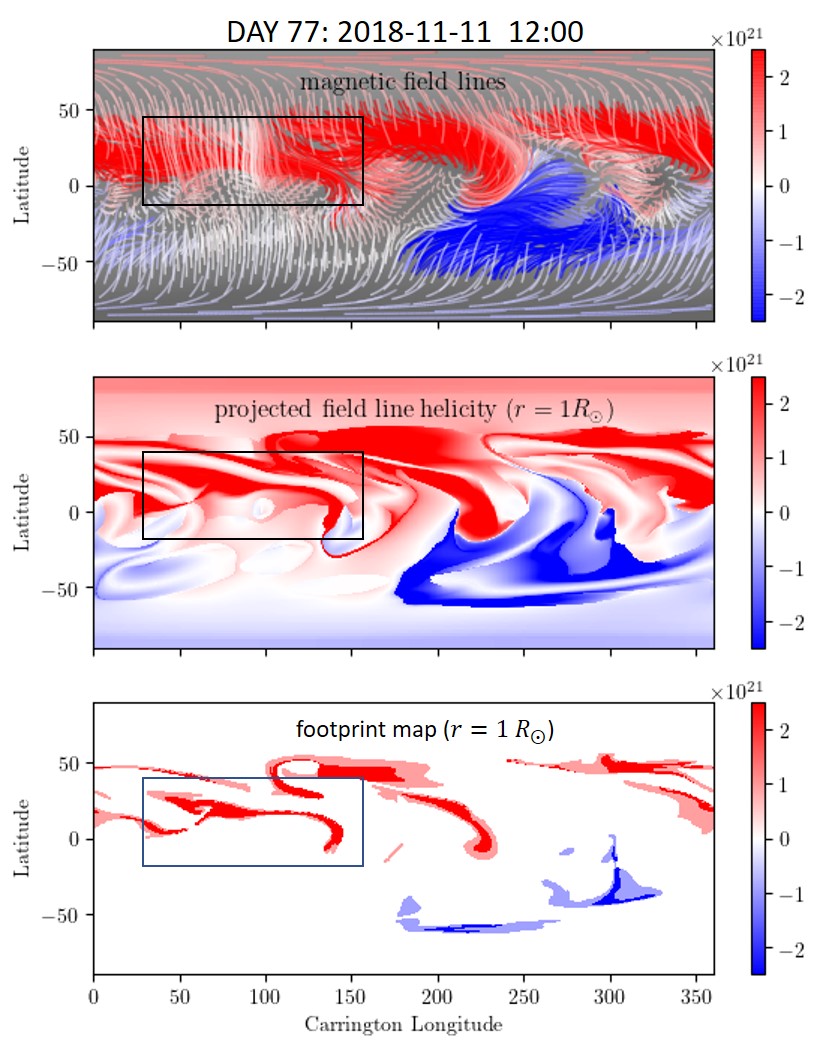}
              }
     \vspace{-0.35\textwidth}   
     \centerline{\Large \bf     
         \hfill}
     \vspace{0.31\textwidth}    
   \centerline{\hspace*{0.015\textwidth}
               \includegraphics[width=0.5\textwidth,clip=]{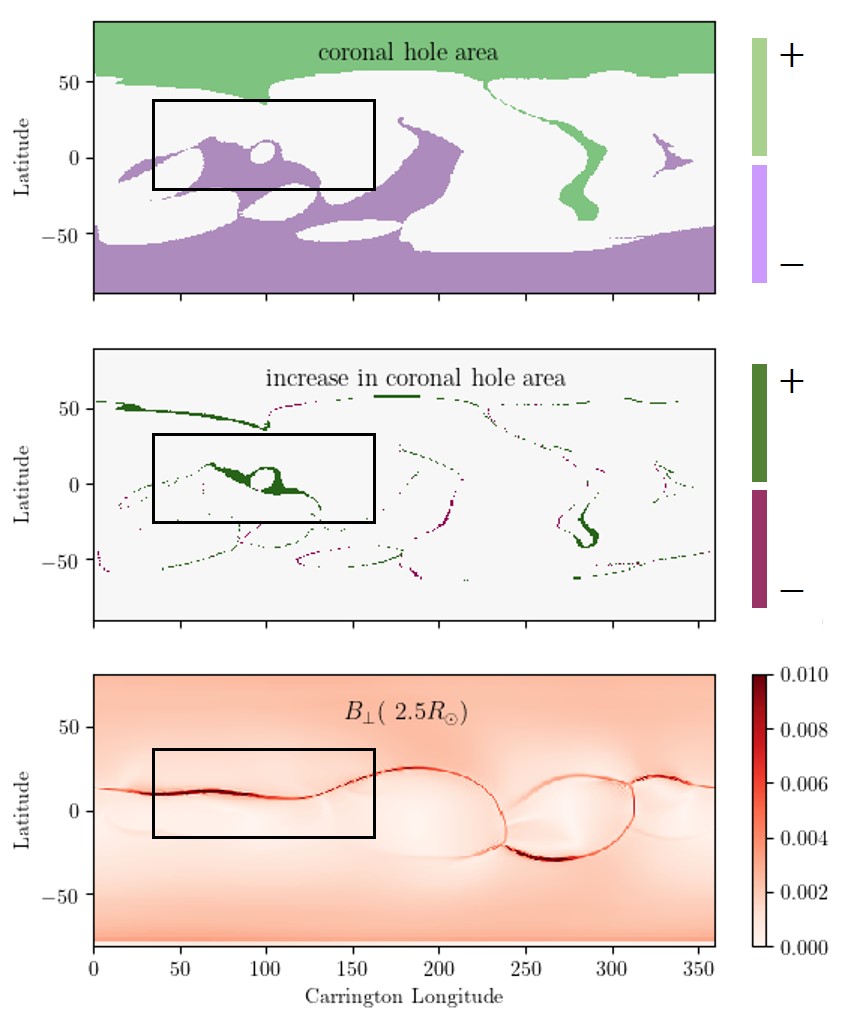}
               \hspace*{-0.03\textwidth}
               \includegraphics[width=0.5\textwidth,clip=]{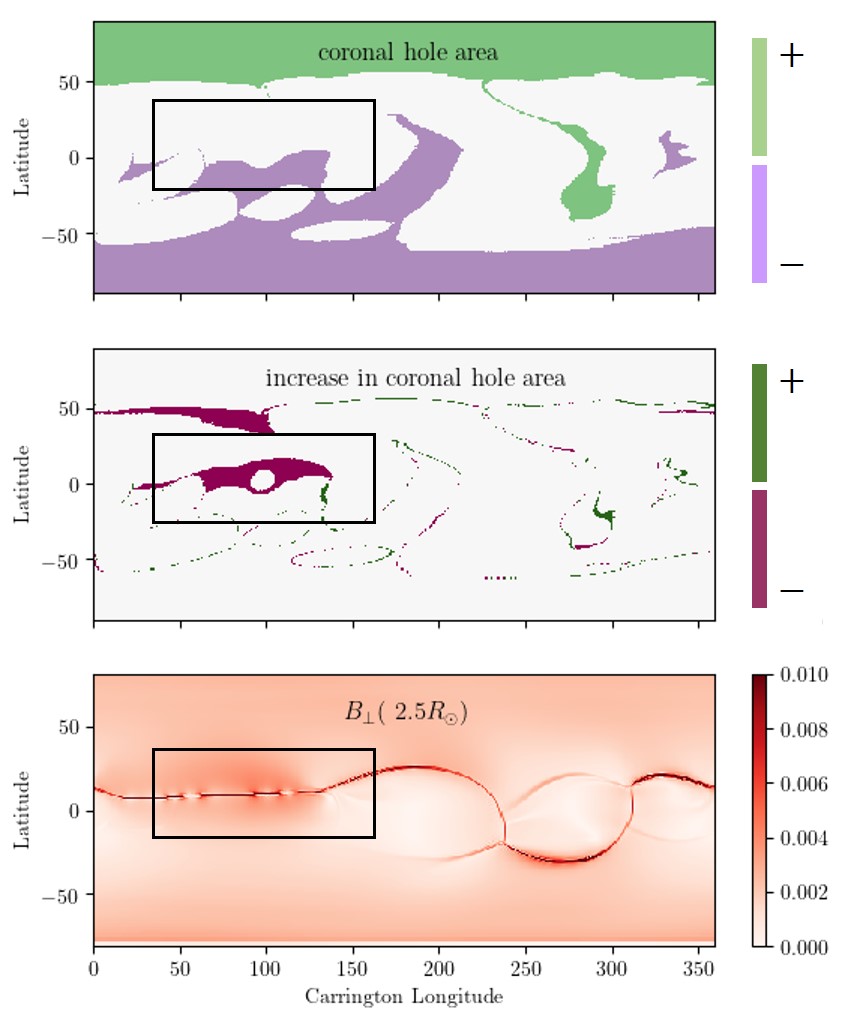}
              }
     \vspace{-0.35\textwidth}   
     \centerline{\Large \bf     
         \hfill}
     \vspace{0.32\textwidth}    
              
\caption{Spatial distributions on two different dates corresponding to the second case study: an overlying-arcade eruption event on Day 76 (10 November 2018). The first row: a top view of the magnetic-field lines colour-coded according to their field-line helicity (positive in red and negative in blue, units in Mx). The grey background refers to the radial component of the photospheric magnetic field (within $\pm$ 10\,G). The second row: photospheric mapping of field-line helicity. The third row: footprint of selected non-potential structures, marking cores and their extension with darker and lighter shades, respectively. The forth row: foot-points of the open-field lines with upward (green) and downward (violet) directions. The fifth row: change in coronal-hole area compared to the previous day; dark green and dark purple suggest increase and decrease in coronal-hole area, respectively. The distribution of the horizontal component of the magnetic field [G] at the outer boundary is depicted in the last row.}
\label{fig6}
\end{figure}

In the first three rows of Figure \ref{fig6}, we present field-line helicity maps on 8 and 11 November 2018 (Days 74 and 77) generated in the same way as Figure \ref{fig2}. A notable change in field-line distribution is visible on the second date (within the rectangular box in the northern hemisphere). Likewise, in the third row, we notice the footprint area of the structure under consideration has shrunk simultaneously. Moreover, the fifth row in Figure \ref{fig6} shows an increase in the coronal-hole area in the initial stage of the event followed by a significant decrease on 11 November 2018, indicating opening up and closing down of magnetic-field lines, respectively. Additionally, in the last row of Figure \ref{fig6} we observe expected signatures in the $B_{\perp}(r = 2.5\,\rm{R_{\odot}})$ maps quite similar to the full flux-rope eruption case. All these changes are reflected in the overall temporal evolution of different measures of non-potentiality as depicted in Figures \ref{fig1} on Day 76 (refer to the solid-magenta-vertical line).   

\begin{figure}[ht!]    
\centerline{\includegraphics[width=0.6\textwidth,clip=]{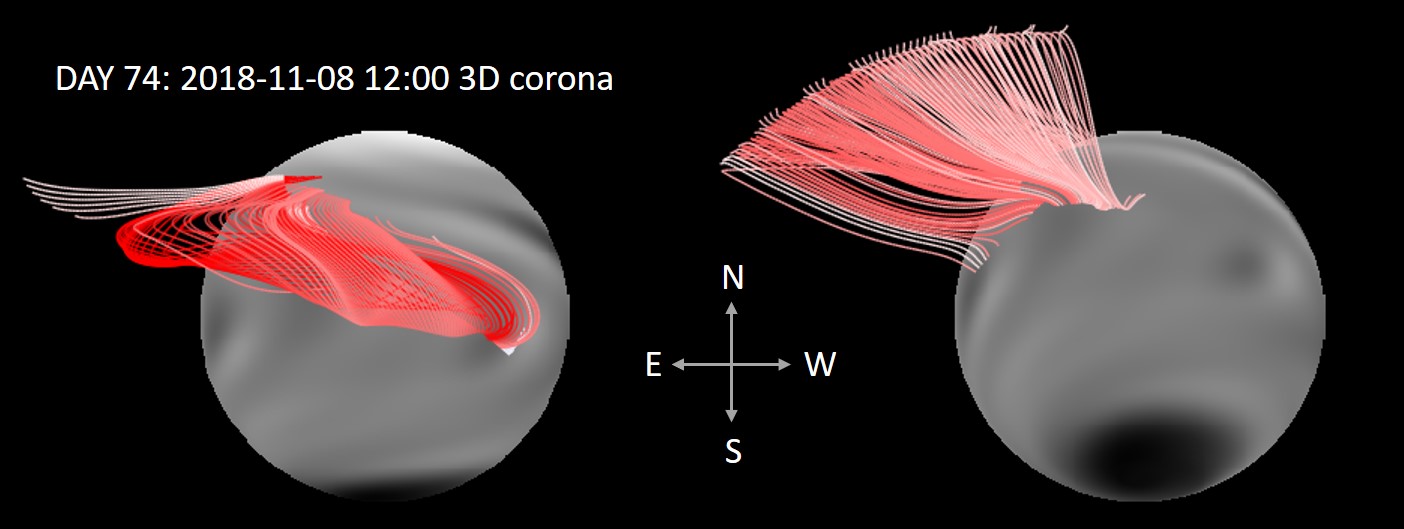}}\vspace{-0.02\textwidth}
\centerline{\includegraphics[width=0.6\textwidth,clip=]{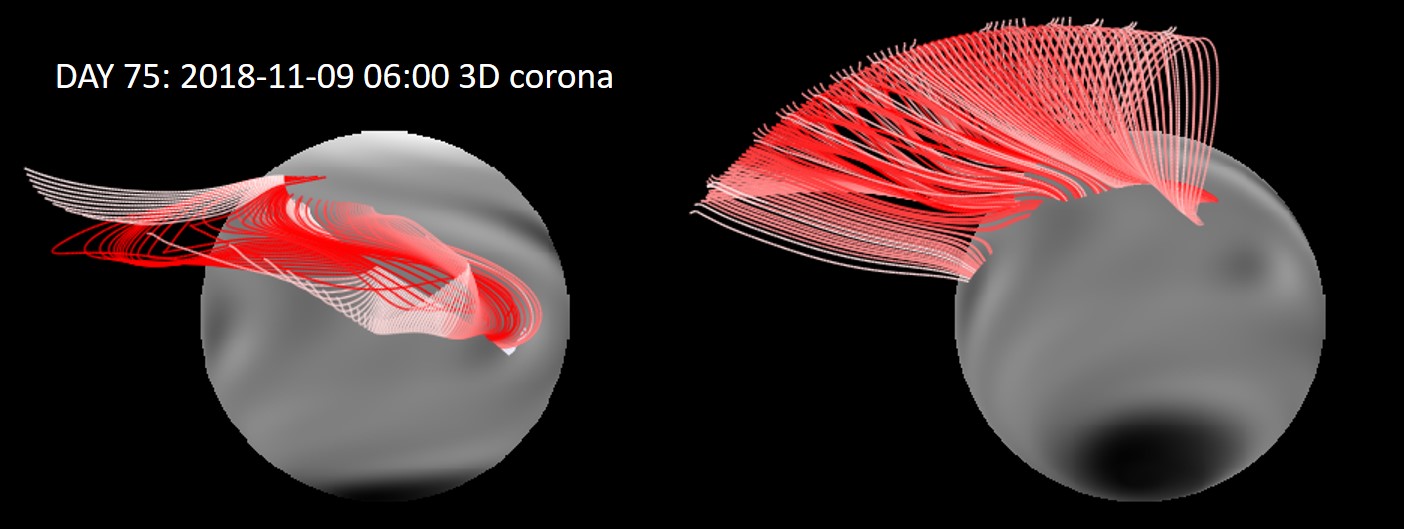}}\vspace{-0.02\textwidth}
\centerline{\includegraphics[width=0.6\textwidth,clip=]{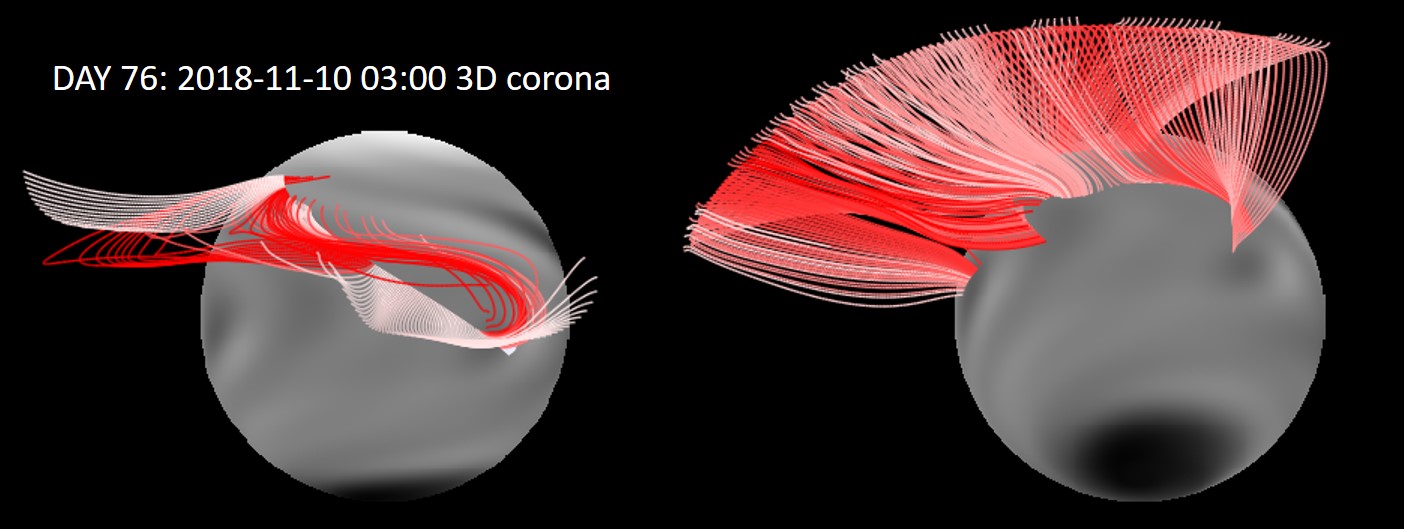}}\vspace{-0.02\textwidth}
\centerline{\includegraphics[width=0.6\textwidth,clip=]{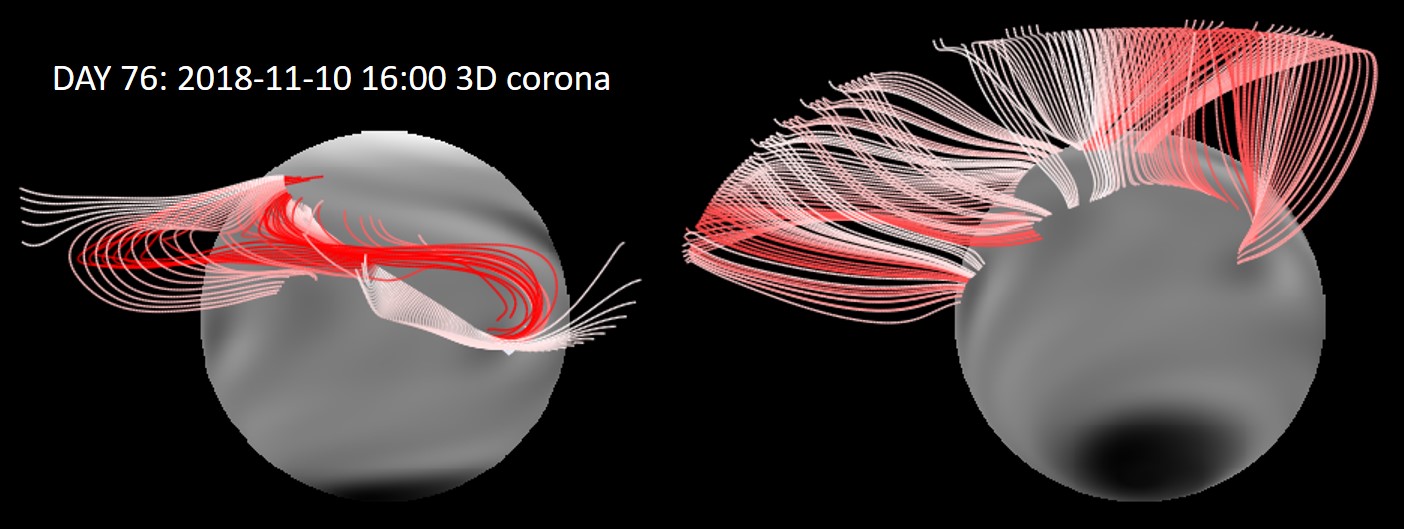}}\vspace{-0.02\textwidth}
\centerline{\includegraphics[width=0.6\textwidth,clip=]{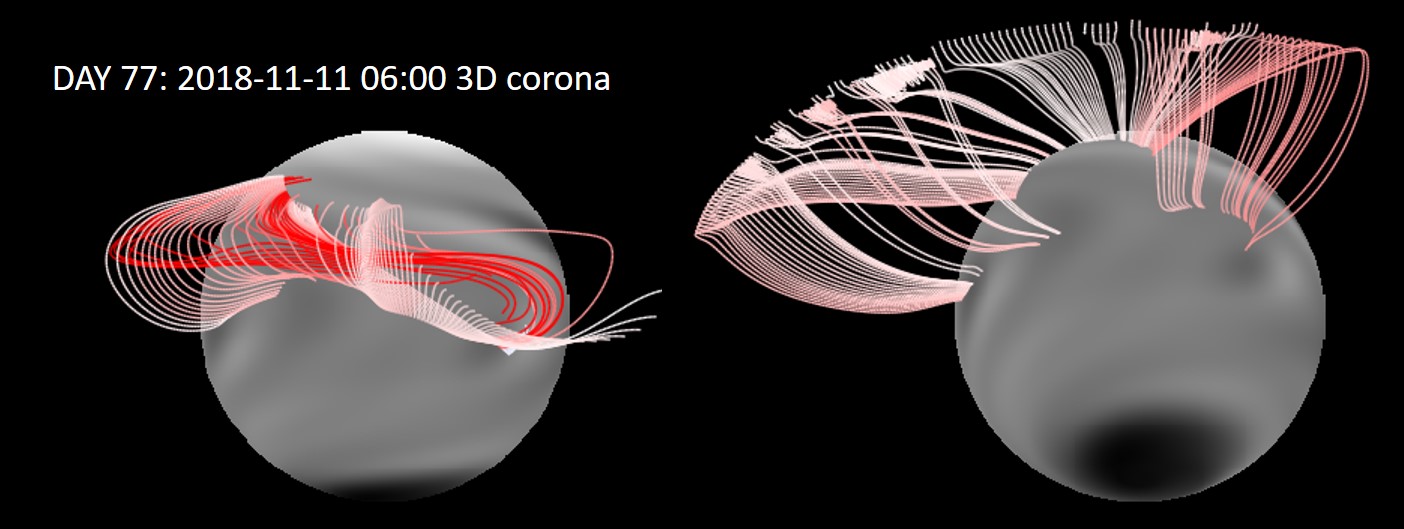}}\vspace{-0.02\textwidth} 
\centerline{\includegraphics[width=0.6\textwidth,clip=]{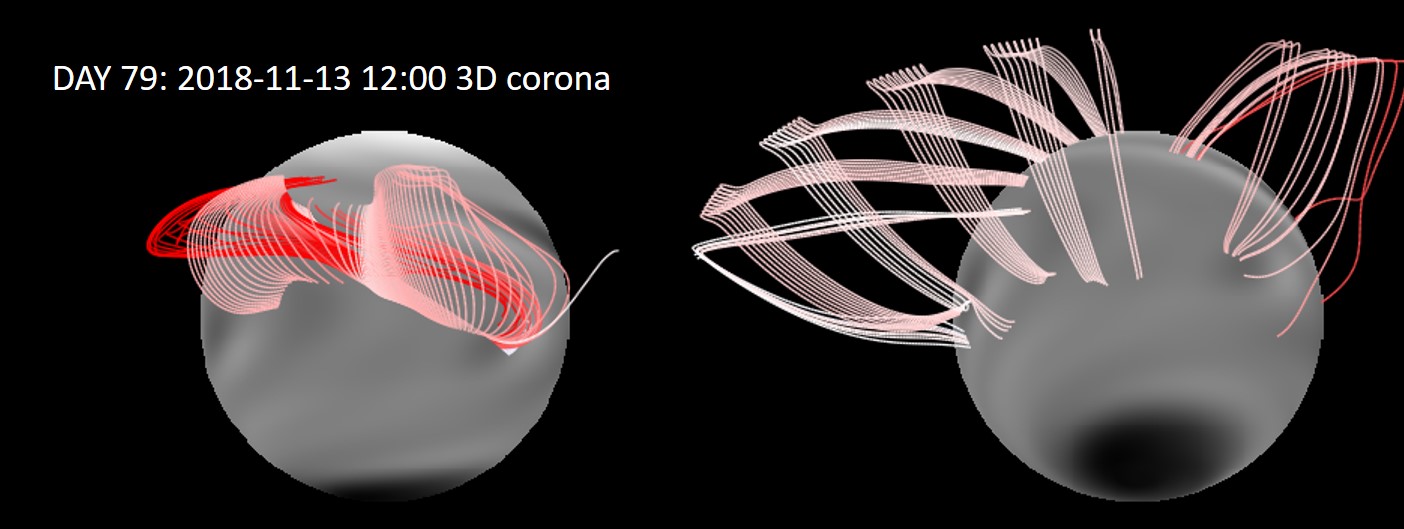}}
\caption{Evolution of the second case study (an overlying-arcade eruption) viewed from two different viewing angles, where the field lines are colour-coded according to the field-line helicity with the maximum amplitude $4.5 \times 10^{21}$\,Mx. The colour red represents positive field-line helicity with the darker shades corresponding to increasing amplitude. The magnetic-field distribution on the solar disk is represented in shades of grey (within $\pm 5$\,G). Left images show field lines traced from the footprint maps, and right images those traced from the outer boundary.}
\label{fig7}
\end{figure}

However, the hourly cadence snapshots during this event reveal quite different dynamics of field lines associated with the non-potential structure, as presented in Figure \ref{fig7}. Similar to the previous case study, field lines shown in the left column are chosen based on the footprint maps. Whereas on the right, field lines passing through the outer boundary with $B_{\perp}(r = 2.5\,\rm{R_{\odot}}) > 0.01$\,G are shown. 

The first salient difference from the previous event is the absence of a clear helical structure like a flux rope in the low-coronal non-potential structure detected by the field-line helicity threshold. Instead, we find highly sheared magnetic-field lines with strong field-line helicity extended in the east--west direction. In the first figure on the top row of Figure \ref{fig7}, we also observe some overlying arcades with lesser but significant field-line helicity encompassing the low-lying highly sheared field lines. As time advances, these overlying arcades undergo reconnection with the surrounding field lines and open up while losing field-line helicity. The same is reflected in the expansion of the coronal-hole area on the fourth row of Figure \ref{fig6}. During the initial stages of this dynamic evolution, we also notice an increase in the magnetic flux through the outer boundary (see the figures on the right side of the first three rows of Figure \ref{fig7}). In the last row, we find some overlying arcades that are almost potential with negligible helicity starting to close down on the existing east--west directed low-lying field lines with much stronger helicity. On the following days, the same process continues (see the last two rows in Figure \ref{fig7}), and the overlying arcade relaxes down by 13 November (Day 79) to a more potential-like structure containing less helicity than the initial arcade on 8 November (Day 74). There is a lesser open field in the lower energy arcade, manifested through the decrease in the coronal-hole area (Figure 6, fourth row). However, unlike unlike the flux-rope eruption event analysed in Section \ref{sec3.2.1}, the low-lying field lines constituting the core of the structure have not changed significantly and are still present. It is the overlying arcade that has erupted, not the sheared field lines within the arcade. These remain almost unaltered throughout the whole event.                    

\begin{figure}[ht!]    
\centerline{\includegraphics[width=1.0\textwidth,clip=]{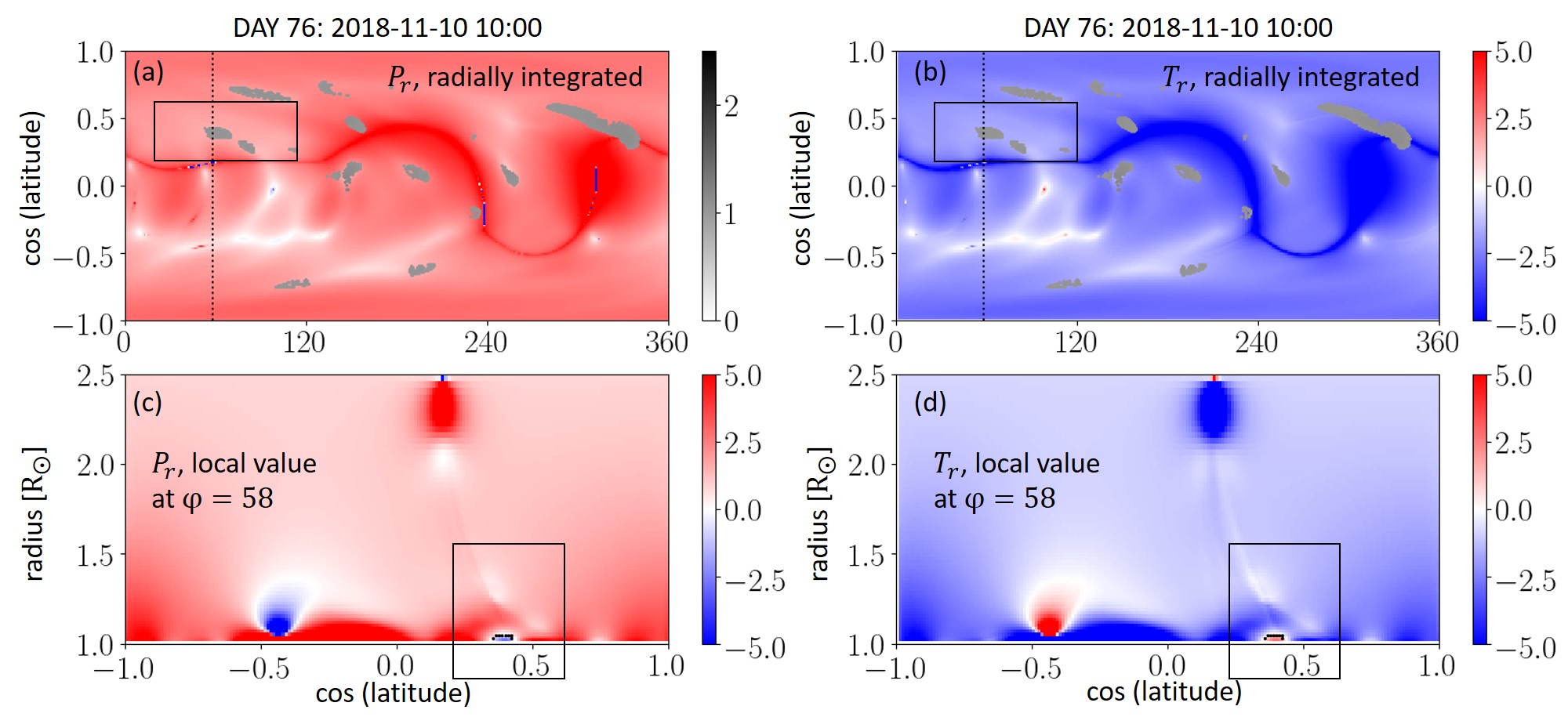}}
\caption{Pressure-gradient force and the radial component of magnetic-tension force for the second case study. Same as in Figure \ref{fig4}, the forces are shown by the red/blue colour map. In (a) and (b), these are coloured by radius according to the black/white colour scale while the viewing angle is perpendicular to the surface. In (c) and (d), the forces are plotted as functions of radius and cos(latitude) across the dotted vertical cut at $58^{\circ}$ longitude.}
\label{fig8}
\end{figure}

\begin{figure}[ht!]    
\centerline{\includegraphics[width=1.0\textwidth,clip=]{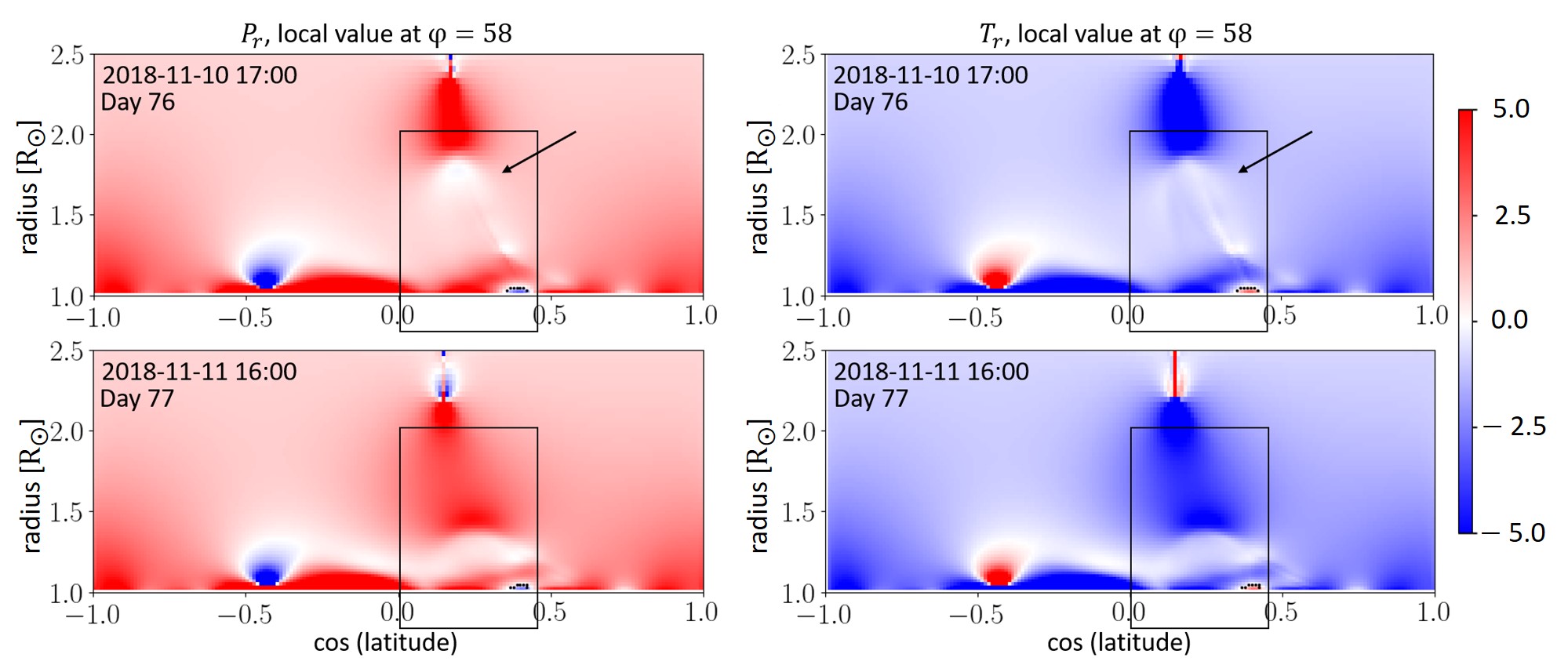}}
\caption{Next stages of evolution of pressure-gradient and radial-tension force across the non-potential structure shown in Figure \ref{fig8} (at $58^{\circ}$ longitude). The values are according to colour bar shown on the right.}
\label{fig9}
\end{figure}

The evolution of radial magnetic-tension force and pressure gradient force cross this structure also shows different behaviour than the case with flux-rope eruption. Its hourly evolution suggested that the non-potential structure did not erupt; rather, field lines in the overlying arcade shed helicity through reconnection and are redistributed to a comparatively stable configuration. In the pressure-gradient and tension-force maps in Figure \ref{fig8}, we notice that there is a structure satisfying the criteria according to Equation [\ref{eq8}], but is situated close to the photosphere on 10 November 2018 (Day 76) at 10.00. The size of the structure is notably small compared to the previous one. Moreover, its evolution is completely different, as depicted in Figure \ref{fig9}. We notice that a new structure that is initially visible in the upper corona (indicated by black dots and pointed by a black arrow) appears and moves downwards while the original structure stays rooted at the bottom layers. At the end of the evolution, the downward-moving structure disappears, but the low-lying part remains almost the same. Clearly, the nature of evolution suggests a redistribution of magnetic structures over the original sheared structure rather than a full-scale eruption of a flux rope. Additionally, from the map sequence of tension force, we do not find any positive values of $T_r (\theta, \phi)$ at 2.0\,$\rm{R_{\odot}}$. Consequently, on Day 76 (indicated by the solid-magenta line in Figure \ref{fig1}), there is no peak visible in the maximum positive $T_r (2.0\,\rm{R_{\odot}})$ evolution. This analysis, thereby, further supports our understanding that this particular event is not associated with the ejection of a pre-existing helical magnetic structure or flux rope.  

We note that, there is a small peak in the maximum positive $T_r (2.0\,\rm{R_{\odot}}, \theta, \phi)$ on Day 122 (see the black dots in the third row of Figure \ref{fig1}). It may seem to be associated with an overlying-arcade eruption occurring on Day 119, as indicated by the dashed-magenta line. However, a careful investigation showed that the peak was linked with the eruption of a small flux rope on Day 122, which was too weak to satisfy the threshold condition of $B_{\perp}$. Thus we do not notice any corresponding peak in the evolution of the strength of $B_{\perp}$ (see the red curve in the second row of Figure \ref{fig1}).

\subsection{Statistical Properties} 
\label{sec3.3}

We performed similar detailed analyses for the remaining 17 events and found them distributed between two classes: full-scale flux-rope eruptions and overlying-arcade eruptions. Based on this classification, these events are separately identified with a set of magenta (arcade eruption) and cyan lines (flux-rope eruption) in Figure \ref{fig1}. Note that we classify an event as a case of overlying-arcade eruption when we cannot find helical magnetic field associated with any existing underlying well-defined flux-rope structure. Rather, these non-potential structures always had highly sheared underlying arcades that remained almost unaltered throughout the evolution of the overlying arcades. It is noteworthy that among the cases with flux-rope eruptions, we encounter two events where the flux rope started forming high up in the corona just before becoming unstable and resulting in a full-scale eruption. In other cases, we see the rise and ejection of preexisting flux ropes through the outer boundary. To go beyond the qualitative distinction in Section \ref{sec3.2}, we seek quantitative measures to differentiate between the two types of event: flux-rope eruption or overlying-arcade eruption. Although our sample of 19 events is small, the statistical analyses can shed more light on these different classes of events and this will be useful in future studies also.

As discussed in Section \ref{sec3.1.2}, each event is accompanied by either partial or complete removal of pre-eruption helicity during its course of evolution, which is clearly visible in the foot-point maps (third row in Figure \ref{fig2}). For each of the 19 events, we select the respective non-potential structure based on a field-line helicity threshold and evaluate three quantities (calculated based on the footprint maps for, e.g., the third row Figure \ref{fig2}): total area, total magnetic flux, and total field-line helicity content. These quantities are measured at the beginning and at the end of each event, such that we can also quantify the changes in those quantities. In general, the most dynamic part of evolution causing disappearing structure happens quite rapidly within one day. Thus to evaluate the changes in area, magnetic
flux, and helicity content, we compare the day before and after the event. The results are summarised in Table \ref{table1}, where we have classified each event into one of the two categories through manual inspection of the hourly cadence magnetic-field lines.

The majority of the non-potential structures (14 out of 19) have footprint area less than $5 \times 10^{20}$\,cm$^{2}$. Various observational studies (\citeauthor{2014LRSP...11....1P}, \citeyear{2014LRSP...11....1P}, and reference therein) found the length of filaments varies from about $3 \times 10^{9}$ to about $1.1 \times 10^{10}$\,cm, and the width ranging between a few $10^{7}$ up to a few $10^{9}$\,cm in different filaments. Based on these statistics, the maximum area of a filament would be a few $10^{19}$\,cm$^{2}$, which is one order lesser than the size of the simulated flux ropes. We speculate that the difference is caused by two factors: firstly, in this simulation, most of the flux ropes are large structures forming over large-scale neutral lines, more akin to polar-crown filaments than to those within active regions. Secondly, dense filaments themselves sit within wider filament channels of sheared magnetic field \citep{2010SSRv..151..333M}, and it is the latter that would correspond to the flux ropes considered here. For the two different types of events, full-scale eruption and overlying-arcade eruption, we did not find any significant difference in their respective footprint area, (see the first row in Table \ref{table1}).

However, when we measure the unsigned magnetic flux associated with these structures on the surface at the beginning of the events, there exists a significant distinction between the two classes of events. The average total magnetic flux for a flux-rope eruption is almost three times higher than that for an arcade eruption event, and the values are $4.83 \times 10^{20}$\,Mx and $1.66 \times 10^{20}$\,Mx, respectively. It is hard to evaluate how these values agree with observation since we did not find a published study on observed magnetic clouds covering the same period as our simulation. However, we can compare our results with those presented by \cite{2017ApJ...846..106L}. They used a similar magnetofriction simulation covering a period spanning from June 1996 to February 2014. Their reported mean unsigned magnetic flux for erupting flux ropes was $4.04 (\pm 6.17) \times 10^{21}$\,Mx -- thus their mean is about one order of magnitude higher than our calculated average. Note that their study included the whole of Solar Cycle 23 and the major portion of Cycle 24 with observed sunspots. In contrast, our study focuses on an interval very close to Cycle 24 minimum and does not include sunspot emergence. Thus, in our simulations, the overall average magnetic-field strength on the surface is relatively small even though the same magnetofriction model has been used. Comparing the unsigned magnetic field associated with the structures, we find the averages are $0.77$\,G and $0.20$\,G for the flux-rope eruption and overlying-arcade eruption events, respectively; thus, the difference persists between these two classes.

The total helicity contained within a non-potential structure has been computed from the field-line-helicity map on the photospheric boundary using Equation [6] of \cite{2017ApJ...846..106L}. We find the average helicity content of the erupting non-potential structures is $3.17 \times 10^{42}$\,Mx$^2$, which is one order less than the value reported by \cite{2017ApJ...846..106L}. Again, this is due to our choice of a significantly less active period near cycle minimum, whereas their average was over the whole solar cycle. Nonetheless, the average is twice as high for the flux-rope eruption events compared to the overlying-arcade eruption events. This difference remains even after considering the area-normalised helicity magnitude (normalisation is based on associated area). If we calculate the average helicity-ejection rate per day over our simulation, we find $2.32 \times 10^{41}$\,Mx$^2$\,day$^{-1}$. This compares well with the (approximately) $3 \times 10^{41}$\,Mx$^2$\,day$^{-1}$ during the (previous) solar minimum seen in the \citet{2017ApJ...846..106L} simulation (see their Figure 13). The latter simulation produced ejection rates of magnetic helicity and flux over the solar cycle that were in agreement with observational estimates from magnetic clouds \citep{2016SoPh..291..531D}, so we have reason to believe that the helicity ejection in our simulation is also broadly consistent with observations.

The distinction between the two classes of eruptions becomes more prominent when evaluating the change in magnetic flux and helicity magnitude. The average changes in magnetic flux on the surface, and total-helicity magnitude are significantly higher in the case of flux-rope eruption events (see Table \ref{table1}). The contrast between these categories remains the same, even when we consider the change in the average magnetic field or normalised helicity magnitude.  

\begin{table}
\caption{Comparison between two classes of events: hyperdiffusion}
\label{table1}
\begin{tabular}{lcc}     
  \hline                   
Quantity & Flux-rope eruption & Overlying-arcade eruption\\
  \hline
number of events & 8 & 11 \\
footprint area [$10^{20}$\,cm$^2$] & 7.03 & 7.42 \\
total magnetic flux [$10^{20}$\,Mx] & 4.83 & 1.66 \\
average magnetic field [G] & 0.77 & 0.20 \\
total helicity magnitude [$10^{42}$\,Mx$^2$] & 3.17 & 1.49 \\
normalised helicity magnitude [$10^{21}$\,Mx] & 3.42 & 1.83 \\
change in magnetic flux [$10^{20}$\,Mx] & 1.44 & 0.12 \\
change in helicity magnitude [$10^{41}$\,Mx$^2$] & 9.23 & 3.28 \\
helicity flux [$10^{42}$\,Mx$^2$] & 5.08 & 2.33 \\
  \hline
\end{tabular}
\end{table}

Finally, we estimate the helicity flux through the outer boundary of the corona during each of the 19 events (from the last row in Figure \ref{fig1}). Note that, unlike the calculated helicity magnitude associated with a non-potential structure, helicity flux is a global quantity. So the estimated value will have contributions from all coronal structures with positive as well as negative helicity magnitude and can add up to a negligible helicity flux. Even so, we find, for flux-rope eruption events, the average magnitude of helicity flux is about two times higher ($5.08 \times 10^{42}$\,Mx$^2$) than for arcade eruption events ($2.33 \times 10^{42}$\,Mx$^2$). Moreover, assuming the helicity flux through the outer boundary to originate primarily from the changing helicity content of the particular structure in consideration (at that same epoch), we further perform a linear correlation analysis between helicity magnitude and respective helicity flux. The evaluated Pearson linear correlation coefficient is $0.71$ (with p-value $0.001$). Interestingly, when the same analysis is performed separately for two distinct classes of events, we find an even higher correlation (coefficient $= 0.85$) between the change in helicity magnitude and the respective helicity flux for flux-rope eruptive events. However, the correlation decreases remarkably for cases with overlying-arcade eruption (coefficient $= 0.47$). This lower value indicates, for the second type of evolution, that the helicity flux is linked with the overlying sheared magnetic-field lines exterior to the selected non-potential structure, whose helicity remains largely unchanged.       

\subsection{Parameter Dependence}
\label{sec3.4}

It is only logical to examine whether the existence of two distinct classes of events is an artefact of the way that we model the non-ideal part in the induction Equation [\ref{eq1}] or any other model parameters. Thus we perform two additional sets of magnetofriction simulations of 180 days starting with the same initial surface magnetic-field map. In one set, we consider ohmic diffusion to model $\vec{N}$ (instead of hyperdiffusion), and in the other we use a slower solar wind (maximum speed 50\,km\,s$^{-1}$). We again evaluate different measures of non-potentiality over 180 days, similar to our previous analyses with hyperdiffusion. 

For both cases, we find a similar increasing trend in non-potentiality with episodic decreases, which are linked to the significant reshaping of coronal magnetic field associated with non-potential structures. A corresponding figure depicting a comparison of the mean field-line helicity among different simulations is provided in the Appendix. We found that,
for about the initial 90 days, the epochs of drastic changes in mean field-line helicity are happening concurrently among the different sets of magnetofriction simulations with hyperdiffusion, ohmic diffusion, or higher outflow speed. We verified that these simultaneous drops do originate from non-potential structures at the same locations in the corona. After 90 days, the simulations with ohmic diffusivity or slower outflow start to diverge notably, such that the temporal and spatial changes in the non-potentiality associated with the structures no longer match with those occurring in the hyperdiffusion simulation.

We identify 17 and 15 events from the simulations with ohmic diffusion and slower outflow, respectively, while following the same analyses described in Section \ref{sec3.1} and then perform additional simulations with hourly cadence for individual events (as in Section \ref{sec3.2}). The slightly later eruptions (in the initial 90 days) and lower number of events overall likely result from the dissipation of stored helicity by the ohmic-diffusion term, in contrast to hyperdiffusion where there is no volume dissipation of helicity. The slower outflow has the effect of allowing the corona to store more helicity, thus delaying eruptions. We found about 41\,$\%$ and 47\,$\%$ out of 17 and 15 events (corresponding to ohmic diffusion and slower outflow, respectively) were related to complete eruption of flux ropes, while the rest were overlying-arcade eruption events. 

We again perform the same statistical analysis (as in Section \ref{sec3.3}) for these two new sets of simulations. The distinction in the amplitude of the associated magnetic flux and helicity magnitude and their respective changes persist for the events with full-scale flux-rope eruption and overlying-arcade eruption. The corresponding tables (\ref{table2} and \ref{table3}) are provided in the Appendix. 

To summarise, altering different elements in the magnetofriction simulation has a finite effect on how the non-potentiality will build up in the coronal magnetic field. However, the existence of two disparate class of evolving magnetic field associated with non-potential structures is independent of our choices of model parameters.      

\section{Concluding Discussion}
\label{sec4}
This work investigates the evolution of coronal magnetic field of the full Sun over six months -- a period chosen very close to the Solar Cycle 24 minimum (26 August 2018\,--\,22 February 2019). Although we did not include any sunspot emergence in the magnetofriction simulation, the slow but continuous building up of non-potentiality during this period leads to a moderately dynamic solar corona. The shearing motion associated with the differential rotation on the solar surface along with magnetic reconnection plays the most vital role in generating complexities in coronal magnetic-field distribution. These complexities are often localised, forming non-potential structures. These structures eventually become unstable and instigate abrupt and drastic changes in the surrounding magnetic field. Through such changes, the coronal magnetic field sheds some of its non-potentiality associated with these structures. Thus we notice sudden decreases in free energy and mean field-line helicity at the same epochs, which we label as individual events. 

Each event is driven by the destabilisation of a non-potential arcade whose photospheric footprint is seen in field-line helicity maps. As the structure becomes unstable, it is pushed through the corona, causing opening up of overlying field lines. Simultaneously we observe significant changes in coronal-hole area and the horizontal component of the magnetic field in the vicinity of the original structure. The eruption of a structure with high field-line helicity also causes notable variation in the open magnetic flux and helicity flux through the corona's outer boundary.

However, we found that these events comprise two distinct classes, even though their associated signatures of non-potentiality are quite comparable. The difference mainly was perceived when we studied individual events with hourly cadence. In one set of events, a low-lying magnetic flux rope rises and disappears entirely, indicating a full-scale eruption. This could explain the origin of occasional large CMEs during solar minimum \citep{2012LRSP....9....3W,2019SSRv..215...39L}. There exist numerous computational and observational studies on CMEs originating from the eruption of flux ropes \citep{2011LRSP....8....1C,2014LRSP...11....1P}, but in this model, they form and erupt self-consistently without the need for an artificial driver such as a localised sheared flow. We speculate that in our model, the initial magnetic reconnection occurs underneath the flux rope forming a ``U''-shaped loop. As this U-loop moves radially outward, it imparts excessive stress to the flux-rope structure, which in turn causes the overlying field lines to break via tether-cutting. The whole dynamics resembles the class of CME models with the ``tether-straining'' mechanism \citep{2001GMS...125..143K}, although of course our magnetofrictional model is only quasi-static. Nevertheless, \cite{2018JSWSC...8A..26P} have shown that configurations that are unstable in the magnetofrictional model are likely to be unstable in full MHD. Indeed, \cite{1986ApJ...311..451C} showed that force-free equilibria have the same linear stability properties in magnetofriction as in ideal MHD. 

In the other class of events, reconnection with the surrounding magnetic-field lines allows the localised non-potential system to partially shed some of its helicity content before settling down to a relatively more stable structure. However, the original sheared structure at lower height remains almost unaltered during the event. Thus any visible changes in mean field-line helicity or any other global measures of non-potentiality are mainly linked with the evolution of the overlying highly sheared arcades and not with the structure in the lower corona. Absence of positive radial tension force calculated at mid-coronal height (2.0\,$\rm{R_{\odot}}$) also suggests that no helical magnetic structures are associated with these events. Such evolution resembles the observed phenomenon known as streamer blowout \citep{2007ApJ...671..926S}, where a slow CME is associated with the gradual swelling then sudden reconnection of a coronal streamer. In a detailed study with the Large Angle and Spectrometric Coronagraph (\textit{LASCO}) on board the Solar and Heliospheric Observatory (\textit{SOHO}) CME observations during 1996\,--\,2012, assisted by 3D MHD simulations and EUV and coronagraphic observations from the Solar Terrestrial Relations Observatory (\textit{STEREO}) and \textit{SDO}, \cite{2013SoPh..284..179V} investigated how many observed CMEs had flux rope structures. Their statistical results based on Cycle 23 CMEs show that 40\,$\%$ of total 2403 CMEs can be regarded as standard three-part CMEs (including loop CMEs) with flux ropes at their cores. However, a significant number of events (40\,$\%$) did not have any detectable flux rope cavity and were named ``outflow'' CMEs, which can be as large as regular CMEs. These outflow CMEs lack a three-part morphology but are too wide to be categorised as jet CMEs, and the front parts are not sharp enough to be classified as loop CMEs. Although \cite{2013SoPh..284..179V} mentioned that the undecided physical nature of these outflow CMEs could partially be an artefact of the lack of good observational data, their study leaves open the possibility of CMEs without pre-existing flux ropes.

\cite{2018ApJ...861..103V} also performed an extensive study with more than 900 streamer-blowout events in the \textit{LASCO}-C2 observations between 1996 and 2015. They found streamer-blowout events, particularly during cycle minimum, to be associated with large-scale neutral lines, and some without signatures of helical flux ropes. Moreover, using a full-MHD numerical simulation, \cite{2016JGRA..12110677L} modelled the evolution of the slow streamer-blowout event of 1\,–-\,2 June 2008 which was also the origin of a stealth CME. The primary characteristic of stealth CMEs is the virtual absence of any identifiable surface or low-corona signatures indicating that an eruption has occurred. Thus, the second kind of events, in our speculation, could also be manifestations of stealth CMEs in terms of the magnetic field. 

For further observational support for our work, we compare the number of CMEs arising from our simulations of the solar minimum corona to the actually observed one recorded by \textit{SOHO}/\textit{LASCO} \citep{2009EM&P..104..295G} during the period when the events start to occur in our simulation. According to the \textit{LASCO} catalogue of CMEs (\url{cdaw.gsfc.nasa.gov/CME_list}), there were 20 ``poor'' CME events detected by both C2 and C3 between 15 October 2018\,--\,22 February 2019 (which is equivalent to days 50\,--\,180 in our simulation, excluding the initial ramp time). In addition, there were additional 17 CMEs detected either by \textit{LASCO}-C2 or -C3. All of them were marked as ``poor'' events, and for many of them estimating associated mass and kinetic energy was impossible. No halo-CME was observed during this period \citep{2020ApJ...903..118D}. Thus the number of events in our simulation (19) appears comparable with the number of recorded CMEs.

We also performed a quantitative analysis of different measures such as area, magnetic flux, and helicity content of the non-potential structures associated with selected events and tried to compare them with observed interplanetary magnetic-cloud statistics. Although our comparison could not be a direct one due to lack of available observations, we were able to validate our range of values -- especially for rates of erupting magnetic flux and magnetic helicity -- by drawing a comparison with another similar long-term simulation of coronal magnetic field performed by \cite{2017ApJ...846..106L}, whose results matched well with observed estimates of magnetic clouds. Thereby, we anticipate that our simulated values would also be close to the observed quantities associated with magnetic clouds during Cycle 24 minimum. 

Finally, to investigate the robustness of our findings, we changed the non-ideal term from hyperdiffusion to ohmic diffusion and changed the solar-wind speed. In both scenarios, we obtain the two distinct classes of eruptive events. From a comparative point of view, eruptions of flux ropes are less frequent than overlying-arcade eruptions in all of the simulations. We think that this is consistent with observational findings that most slow CMEs are not associated with erupting filaments \citep{2008JGRA..113.1104H}.

In conclusion, our study demonstrates that during a low-activity phase of the sunspot cycle, the large-scale shearing plasma flow on the surface can lead to a slow but significant generation of non-potentiality in the coronal magnetic field. Highly sheared arcades and flux ropes with strong field-line helicity store this excess non-potential energy, which is eventually released to the heliosphere through two different classes of eruptive events. This work based on magnetofrictional simulations that exclude any intricacies of emerging sunspots can thus be perceived as the first step to understand the complex global evolution of the coronal magnetic field.

\begin{acks}
This work was supported by STFC (UK) consortium grant ST/S000321/1. The \textit{SDO} data are courtesy of NASA and the \textit{SDO}/\textit{HMI} science team. The CME catalogue is generated and maintained at the CDAW Data Center by NASA and The Catholic University of America in cooperation with the Naval Research Laboratory. \textit{SOHO} is a project of international cooperation between ESA and NASA. We are also thankful to the reviewer for the useful suggestions, which have helped to improve the quality of this manuscript. 
\end{acks}

{\footnotesize\paragraph*{Disclosure of Potential Conflicts of Interest}
The authors declare that they have no conflicts of interest.}

\begin{figure}[!ht]    
\centerline{\includegraphics[width=0.7\textwidth,clip=]{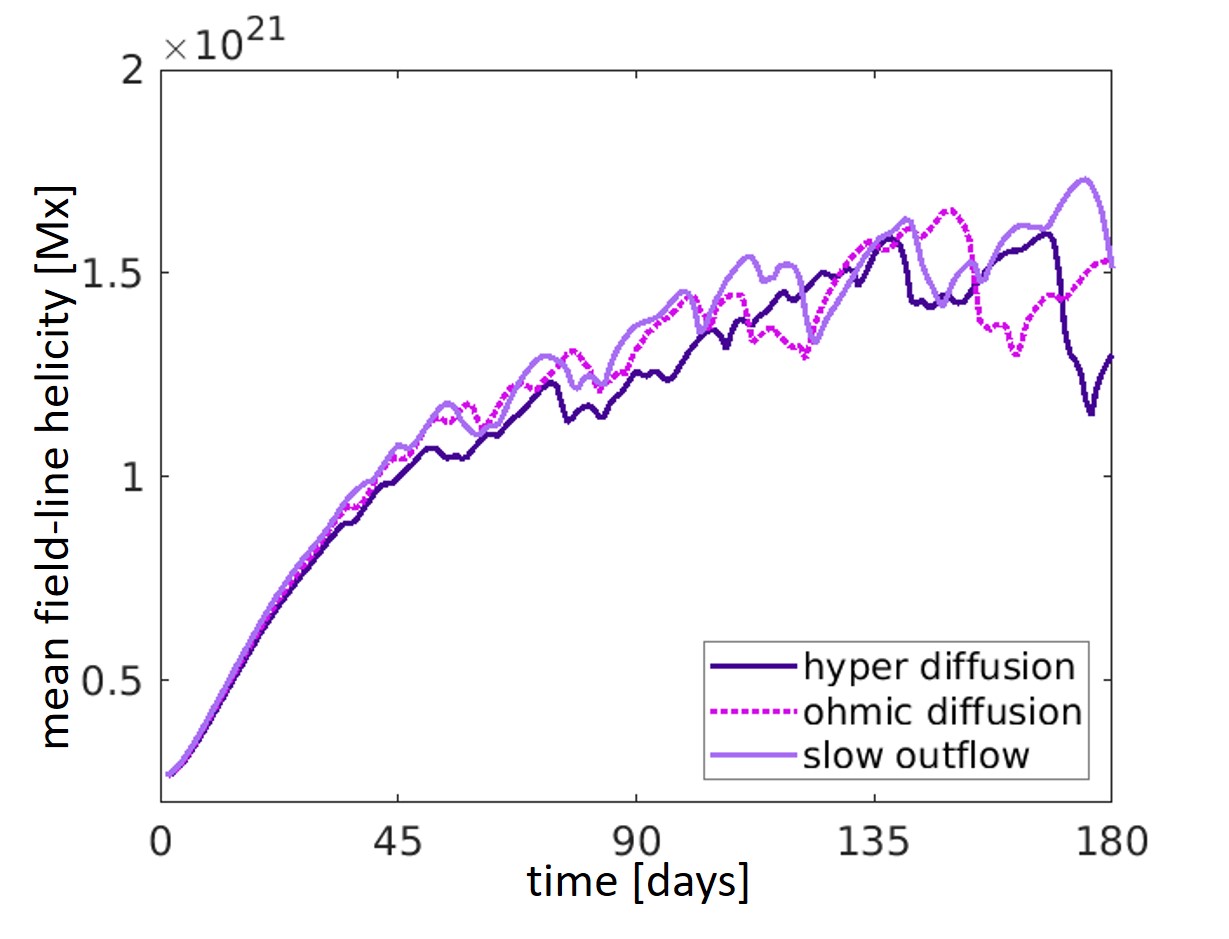}}
\caption{Temporal evolution of mean field-line helicity for different parameter settings.}
\label{fig10}
\end{figure}

\appendix 
Among many measures, we only demonstrate the temporal evolution of mean field-line helicity in Figure \ref{fig10} in the coronal simulations with ohmic diffusion and slower outflow and compare them with the hyperdiffusion case. The overall agreement between the results obtained from ohmic diffusion and hyperdiffusion suggests that both processes generate similar non-potentiality in the coronal magnetic field in general with certain distinctions (discussed in Section \ref{sec3.4}). 

\begin{table}
\caption{Comparison between two classes of events: ohmic diffusion}
\label{table2}
\begin{tabular}{lcc}     
  \hline                   
Quantity & Flux-rope eruption & Overlying-arcade eruption\\
  \hline
number of events & 7 & 10 \\
footprint area [$10^{20}$\,cm$^2$] & 6.47 & 8.25 \\
total magnetic flux [$10^{20}$\,Mx] & 4.33 & 2.61 \\
average magnetic field [G] & 0.67 & 0.28 \\
total helicity magnitude [$10^{42}$\,Mx$^2$] & 2.22 & 1.79 \\
normalised helicity magnitude [$10^{21}$\,Mx] & 3.37 & 1.99 \\
change in magnetic flux [$10^{20}$\,Mx] & 2.26 & 0.73 \\
change in helicity magnitude [$10^{41}$\,Mx$^2$] & 11.53 & 4.18 \\
helicity flux [$10^{42}$\,Mx$^2$] & 4.72 & 3.06 \\
  \hline
\end{tabular}
\end{table}

\begin{table}
\caption{Comparison between two classes of events: slower outflow}
\label{table3}
\begin{tabular}{lcc}     
  \hline                   
Quantity & Flux-rope eruption & Overlying-arcade eruption\\
  \hline
number of events & 7 & 8 \\
footprint area [$10^{20}$\,cm$^2$] & 9.29 & 8.92 \\
total magnetic flux [$10^{20}$\,Mx] & 6.41 & 2.32 \\
average magnetic field [G] & 0.69 & 0.26 \\
total helicity magnitude [$10^{42}$\,Mx$^2$] & 4.20 & 1.12 \\
normalised helicity magnitude [$10^{21}$\,Mx] & 3.88 & 1.79 \\
change in magnetic flux [$10^{20}$\,Mx] & 2.58 & 1.05 \\
change in helicity magnitude [$10^{41}$\,Mx$^2$] & 12.76 & 3.68 \\
helicity flux [$10^{42}$\,Mx$^2$] & 5.09 & 3.13 \\
 \hline
\end{tabular}
\end{table}

We perform additional statistical analyses based on how different physical measures change during the evolution of individual structures associated with the classes of events identified in ohmic diffusion and slower outflow simulations. The results are presented in the following tables \ref{table2} and \ref{table3}. 

\newpage

\bibliographystyle{spr-mp-sola}
\bibliography{solar_bibliography}  

\end{article} 

\end{document}